\documentclass[11pt]{article}

\usepackage[margin=0.85in]{geometry}
\usepackage{amsmath,amssymb,amsfonts,mathtools}
\usepackage{booktabs}
\usepackage{graphicx}
\usepackage{array}
\usepackage{tabularx}
\usepackage{multirow}
\usepackage{enumitem}
\usepackage{algorithm}
\usepackage{algpseudocode}
\usepackage{xurl}
\usepackage{xcolor}
\definecolor{pivotalblue}{HTML}{5F90BD}
\usepackage{microtype}
\usepackage{caption}
\usepackage{float}
\usepackage[section]{placeins}
\usepackage{hyperref}

\hypersetup{colorlinks=true,linkcolor=blue,citecolor=blue,urlcolor=blue}
\graphicspath{{./}}
\setlist[itemize]{topsep=3pt,itemsep=2pt,leftmargin=1.5em}
\setlist[enumerate]{topsep=3pt,itemsep=2pt,leftmargin=1.7em}
\newcommand{\calD}{\mathcal{D}}
\newcommand{\calM}{\mathcal{M}}
\newcommand{\calC}{\mathcal{C}}

\newcommand{\KL}{D_{\mathrm{KL}}}
\newcommand{\indep}{\perp\!\!\!\perp}
\newcommand{\Normal}{\mathcal{N}}
\title{\makebox[\textwidth][c]{\scalebox{0.91}[1]{\fontsize{18}{21}\selectfont\mdseries Privacy-Preserving AI Verification via Minimal Information Disclosure}}}
\author{%
Sleem Abdelghafar\\[-0.2em]
{\small Rice University}
\and
Gabriel Kulp\\[-0.2em]
{\small Intelligence Security Laboratories}
}
\date{}

\begin{document}
\maketitle
\begingroup
\makeatletter
\renewcommand{\@makefnmark}{}
\makeatother
\footnotetext[0]{Correspondence to: Sleem Abdelghafar (\texttt{msm15@rice.edu}).}
\endgroup
\vspace{-1.4em}

\begin{abstract}
AI verification crosses a trust boundary: a verifier must learn enough to establish an authorized claim, yet the same evidence can reveal sensitive details about the model, workload, or hardware. We introduce minimal information disclosure (MID), which designs and quantifies the information content of verifier-facing evidence itself. MID measures collateral leakage with conditional mutual information: what the release reveals about the protected property after the authorized result is known. MID is general by design: it can accommodate different verification goals, protected properties, evidence sources, and deployment constraints. To demonstrate MID’s practicality, we evaluate it on four physical measurements and six verification tasks spanning execution type, hardware identity, compute scale, and model identity. These experiments use three mechanism-design variables—the evidence channel, collection policy, and release transformation—but MID is not limited to these choices and can accommodate other deployable mechanisms.
Across these tasks, MID produces three releases with perfect held-out verification and zero measured collateral leakage, while the remaining tasks yield explicit privacy--utility frontiers. MID also supports ZKP-certified releases: we demonstrate our proposed linear-projection mechanism using a Groth16 zk-SNARK.
\end{abstract}

\section{Introduction}

Suppose a verifier must determine whether a data center trained a model, stayed within a compute limit, used an approved accelerator, or passed an evaluation~\cite{reuel2025open,sastry2024compute,ogara2025hem,baker2025sixlayers,harack2025verification}. The power traces, telemetry records, or electromagnetic measurements that support those claims may also reveal private workload details, model architecture, or hardware configuration~\cite{modelspy,deeptheft,rahman2026hidden,horvath2024sok,gregersen2024power}. Minimal information disclosure (MID) addresses this tension by designing the information content of verification evidence itself: enough to establish the authorized claim while revealing as little as possible about a declared protected property beyond it. Authentication can establish where the evidence came from, and ZKPs can establish how a report was computed; MID determines what the report should reveal. In July 2026, a statement signed by more than 1,200 employees of frontier AI companies called for an international effort to develop the technical and governance tools needed to deliberately pace frontier AI development~\cite{pacingfrontier2026}. Credible evidence of compliance will be central to such coordination, making the design problem addressed by MID increasingly consequential.

\begin{figure}[!t]
\centering
\includegraphics[width=0.98\textwidth]{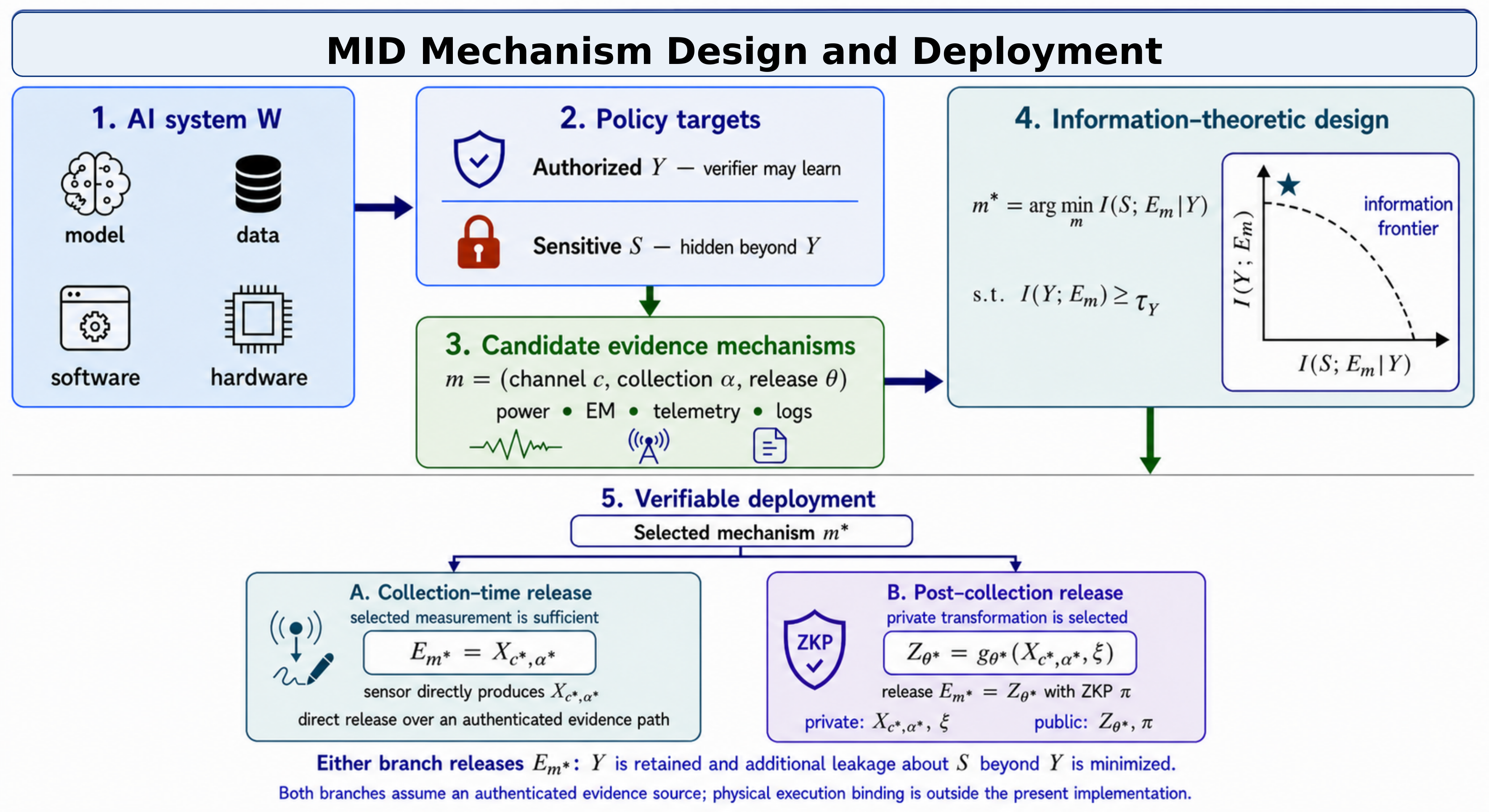}
\caption{MID mechanism design and deployment. For an AI execution, a policy specifies an authorized target $Y$, a sensitive target $S$, and candidate evidence mechanisms $m=(c,\alpha,\theta)$. Information-theoretic optimization selects $m^\star$ to retain authorized information while minimizing additional information about $S$. Deployment then either releases the selected collection-time measurement directly or keeps the measurement private and releases the selected transformed value with a ZKP of correct computation.}
\label{fig:mid-design-deployment}
\end{figure}

\textbf{Why MID?} Sastry et al. suggest that a regulator might receive only a single compliance bit; Baker et al. describe a yes/no determination as the minimal output of a compliance test; and Harack et al. propose one-bit outputs for human assessments conducted in a controlled facility~\cite{sastry2024compute,baker2025sixlayers,harack2025verification}. Other proposals and systems use narrowly specified, minimal, or fixed-alphabet reports~\cite{ogara2025hem,petrie2025flexheg,cankaya2026system,rowstron2026agentic}. These works establish the importance of controlling what reaches the verifier. MID addresses an unresolved design question: among candidate reports, which one reveals the least beyond the authorized answer while preserving verification? Evidence size alone cannot answer this question.

Auditor-in-a-Box provides a concrete example: it presents a deterministic output filter intended to impose a one-bit worst-case disclosure bound, and its published Monitor Query Validation evaluation reports a \textsc{valid}/\textsc{invalid} verdict for each researcher specification~\cite{rinberg2026auditor,penchas2026enabling}. This restricted output limits how much can leave, but does not make the verdict privacy-preserving with respect to a protected property; on the authors' evaluation distribution, the verdict measurably reveals request group.
We focus on requests that human reviewers accepted. The monitor's one-bit decision still distinguishes two sets in the authors' evaluation: one contains many adjacent, proxy, or borderline requests, while the other consists largely of ordinary research requests. A simple predictor learns from the other test cases and then receives only the \textsc{valid}/\textsc{invalid} decision for a new request---not the request itself. It identifies which set the request came from with 72.4\% balanced accuracy, compared with 50\% without the decision. Appendix~\ref{app:auditor-output-privacy} gives the full experiment. Our RL experiment, Figure~\ref{fig:rl-frontier}, shows the same problem for AI-execution evidence: several one-bit reports reveal exact-workload information, whereas MID selects one that achieves perfect held-out RL-versus-non-RL verification with zero measured collateral leakage. MID is not limited to one-bit reports: the same design question applies to richer evidence, motivating MID's general formulation below.

For an AI execution $W$, the policy specifies an authorized target $Y$ that the verifier may learn and a protected property $S$ that should remain hidden beyond what $Y$ necessarily reveals. A candidate mechanism $m=(c,\alpha,\theta)$ produces verifier-facing evidence $E_m$: $c$ selects the evidence channel, $\alpha$ controls how it is collected, and $\theta$ optionally transforms it before release. MID evaluates each mechanism using two quantities. Mutual information $I(Y;E_m)$ measures how informative the evidence is about the authorized target---the information available for verification. Conditional mutual information $I(S;E_m\mid Y)$ measures what the evidence additionally reveals about the protected property after the authorized result is known---the collateral leakage. We formulate the Privacy-Preserving AI Verification Problem as
\begin{equation}
  \min_{m\in\calM} I(S;E_m\mid Y)
  \quad\text{subject to}\quad
  I(Y;E_m)\ge \tau_Y.
  \label{eq:intro-objective}
\end{equation}
\newpage
Equation~\eqref{eq:intro-objective} asks which deployable evidence mechanism reveals the least about $S$ among those sufficiently informative to verify $Y$. The threshold $\tau_Y$ sets the required verification utility, and conditioning on $Y$ prevents MID from penalizing information inherent in the authorized answer. The information content of $E_m$ is therefore the design object; $c$, $\alpha$, and $\theta$ are controls through which MID shapes it. Our experiments use these three controls, but MID is not limited to these controls and can accommodate other deployable mechanisms simply by extending $\calM$. By Fano's inequality, the resulting evidence limits the success of any adversary attempting to infer $S$ from $(E_m,Y)$.

The mechanism variables map directly onto the two deployment paths in Figure~\ref{fig:mid-design-deployment}. When channel and collection choices are sufficient, a collection-time mechanism allows the evidence source itself to produce a low-information measurement that meets the verification requirement; no richer digital trace or private post-processing step is needed. When a post-collection transformation achieves lower leakage at the required utility, the richer measurement remains within a protected boundary and only the selected transformation is released, accompanied by a ZKP of correct computation. Both paths require an authenticated evidence source. Together, they allow MID to accommodate different deployment constraints rather than assume a single architecture. We instantiate these paths using physical evidence channels because they can observe execution independently of the system being checked and because prior side-channel studies provide concrete tests of what such evidence can reveal~\cite{modelspy,deeptheft,horvath2024sok,gregersen2024power}.

To the best of our knowledge, MID is the first AI-verification framework to design the information content of verifier-facing evidence itself by selecting among feasible mechanisms to minimize disclosure about a declared protected property beyond the authorized result while satisfying a verification-utility requirement. Section~\ref{sec:related} distinguishes this contribution from systems that authenticate measurements, protect private inputs, prove prescribed computations, or constrain verifier-facing reports~\cite{ogara2025hem,petrie2025flexheg,harack2025verification,cankaya2026system,waiwitlikhit2024zkaudit}; Section~\ref{sec:discussion} explains how MID can be integrated with these methods.

\noindent\textbf{Our specific contributions are as follows:}
\begin{itemize}
  \setlength{\itemsep}{0.25em}
  \setlength{\parskip}{0pt}
  \item We formulate privacy-preserving AI verification as an information-theoretic optimization problem. MID selects among deployable evidence mechanisms to preserve the information needed for authorized verification while minimizing collateral leakage about a protected property. This makes the information content of verifier-facing evidence itself the design object.
  \item We develop collection-time and post-collection mechanisms that let MID shape evidence through channel selection, collection policy, and release transformation. We show how a selected mechanism can be deployed as an authenticated direct release or as a release computed from private evidence and certified by a ZKP, and demonstrate our linear-projection mechanism with a Groth16 zk-SNARK~\cite{groth2016}.
  \item We demonstrate MID across six physical-measurement tasks spanning execution type, hardware identity, compute scale, and model identity. Three selected mechanisms achieve perfect held-out verification with zero measured collateral leakage, while the remaining tasks yield explicit privacy--utility frontiers. We also evaluate the selected releases against the ModelSpy and DeepTheft attack methods~\cite{modelspy,deeptheft}.
  \item We show, using Auditor-in-a-Box's released evaluation, that even a one-bit verdict can disclose information beyond its intended answer. A predictor that receives only the verdict identifies which of the authors' two request sets a private request came from with 72.4\% balanced accuracy, showing why limiting output size alone is not sufficient.
\end{itemize}

\section{MID Mechanism Design}
\label{sec:method}

MID turns the policy's disclosure boundary into an information-theoretic mechanism-selection problem. We first define the authorized and protected properties, candidate evidence mechanisms, and trust assumptions, then formalize the optimization that selects what the verifier should receive.

\subsection{Problem Setup}
\label{sec:problem}

Privacy-preserving AI verification begins with a policy decision: what may the verifier learn, what must remain protected beyond that answer, and what verification utility is required? Let $W$ denote an execution episode, including any relevant model, data, software, and hardware state. The policy or audit rule defines
\[
Y=f_{\mathrm{ver}}(W),
\qquad
S=f_{\mathrm{sens}}(W).
\]
$Y$ is the authorized target. $S$ is a property that should not be revealed beyond what follows from $Y$, such as model identity, architecture, training data, prompt family, workload purpose, or fine-grained resource use. A policy may specify several sensitive targets $S_1,\ldots,S_J$.

Four roles are conceptually distinct. A policy setter specifies $Y$, the protected properties $S$, and the required verification utility---how informative the evidence must be. An operator executes $W$. An evidence source observes some consequence of that execution. A verifier receives the released evidence and decides whether to accept the authorized claim. One organization may occupy several roles, but separating them distinguishes two security goals: the underlying verification system must establish that the evidence is trustworthy, while MID limits what the verifier can learn from that evidence.

\paragraph{Evidence mechanisms.}

Once the policy fixes $Y$ and $S$, MID designs the evidence passed from the execution to the verifier. During development, a rich baseline observation $R\sim P_R(\cdot\mid W)$ is available to compare and calibrate candidate mechanisms. It can be a high-rate trace, a multichannel telemetry record, a log, a stored model or related record, or a collection of these. The deployed system need not create or retain $R$. A candidate mechanism is
\[
m=(c,\alpha,\theta)\in\calM,
\]
where $c$ selects an evidence channel or channel family, $\alpha$ controls collection, and $\theta$ controls an optional post-collection release transformation. At deployment, the selected sensor and collection rule directly produce
\[
X_{c,\alpha}\sim P_{c,\alpha}(\cdot\mid W),
\]
without first creating $R$. An optional release transformation produces
\[
Z_\theta=g_\theta(X_{c,\alpha},\xi),
\]
where $\xi$ is the release mechanism's randomness. The principal release is
\begin{equation}
E_m^{\mathrm{rel}}=
\begin{cases}
X_{c,\alpha}, & \text{collection-time release},\\
Z_\theta, & \text{post-collection release}.
\end{cases}
\label{eq:mechanism}
\end{equation}
The complete verifier-visible evidence $E_m$ consists of this release and any other fields the protocol sends to the verifier.
Our offline experiments derive $X_{c,\alpha}$ from $R$. In deployment, a collection-time mechanism implements $(c,\alpha)$ at the evidence source, so the richer digital record $R$ need not be created, transmitted, or stored. If a post-collection transformation is selected, $X_{c,\alpha}$ remains inside a sensor, secure meter, or TEE and only $Z_\theta$ is disclosed.

Selecting memory capacity rather than dynamic power, lowering an electromagnetic (EM) sampling rate, and releasing one noisy number are therefore all mechanism choices. MID chooses among them.

Hard decisions enter MID as candidate evidence mechanisms: their source must be bound to the execution, and selection evaluates whether their errors depend on $S$ at fixed $Y$.

\paragraph{Threat model.}

The operator is the integrity adversary: it may attempt to fabricate, omit, replay, or misattribute evidence. The underlying verification system must therefore authenticate the evidence source and bind its output to the relevant execution; MID assumes this integrity layer operates according to its stated security assumptions. It may be implemented by an auditor-controlled sensor, signed secure meter, trusted accelerator firmware, hardware attestation, monitored network boundary, or redundant observations~\cite{cankaya2026system,petrie2025flexheg}. For post-collection mechanisms, a ZKP can bind the released report to the authenticated private measurement.

The verifier is the privacy adversary. It is authorized to learn $Y$ but is not trusted with $S$. It receives $E_m$, knows the selected mechanism, and may use any inference method to recover $S$. MID limits this risk by minimizing $I(S;E_m\mid Y)$, where $E_m$ includes the complete verifier-visible interface, including metadata, commitments, signatures, and ZKP transcripts. MID does not make a compromised evidence source honest or establish that its measurement is truthful; it determines what authenticated evidence should reveal.

MID is calibrated using grouped, labeled executions from the intended environment; the policy declares the protected targets before selection. The mechanism designers may access $R$ during development, while the deployed interface releases only the selected evidence and required verification fields.
\subsection{Information-Theoretic Optimization}

Equation~\eqref{eq:intro-objective} is our information-theoretic formulation of the Privacy-Preserving AI Verification Problem: among deployable evidence mechanisms, select the one that minimizes disclosure about $S$ beyond $Y$ while meeting the required verification utility. The same problem can be written generally by letting $U_Y(E_m)$ denote verification utility and $L_S(E_m\mid Y)$ denote leakage beyond the authorized target:
\begin{equation}
m^\star=\arg\min_{m\in\calM} L_S(E_m\mid Y)
\quad\text{s.t.}\quad U_Y(E_m)\ge \tau_Y.
\label{eq:operational}
\end{equation}
For several sensitive targets, we minimize the largest leakage:
\begin{equation}
m^\star=\arg\min_{m\in\calM}\max_{1\le j\le J}L_{S_j}(E_m\mid Y)
\quad\text{s.t.}\quad U_Y(E_m)\ge\tau_Y.
\label{eq:robust-objective}
\end{equation}
This form controls each declared target separately; if their combination is also sensitive, the vector $(S_1,\ldots,S_J)$ is included as another target. We use mutual information for authorized utility and conditional mutual information for sensitive leakage~\cite{cover2006}:
\begin{equation}
\begin{aligned}
U_Y(E_m)&\triangleq I(Y;E_m),\\
L_S(E_m\mid Y)&\triangleq I(S;E_m\mid Y).
\end{aligned}
\label{eq:cmi}
\end{equation}
The first measures evidence about the authorized target; the second measures additional evidence about $S$ after $Y$ is known. With these definitions, Eq.~\eqref{eq:robust-objective} minimizes $\max_j I(S_j;E_m\mid Y)$. We use base-2 logarithms. Zero leakage is equivalent to $S\indep E_m\mid Y$: the release may reveal $Y$ while conveying nothing further about $S$. If $S$ and $Y$ are correlated, learning the authorized result can still reveal something about $S$; MID treats that as authorized and penalizes only the additional disclosure carried by the evidence.

For discrete $Y$, the best possible point is
\begin{equation}
I(Y;E_m)=H(Y),\qquad I(S;E_m\mid Y)=0.
\label{eq:ideal-point}
\end{equation}
It is possible only if some allowed release determines $Y$ and satisfies $S\indep E_m\mid Y$. At that point, $P(S\mid E_m,Y)=P(S\mid Y)$ almost surely. Otherwise, the allowed mechanisms define the best achievable privacy--utility tradeoffs, which is fully characterized by the information frontier (e.g., Figure~\ref{fig:training-frontier}).

The objective does not depend on a particular attack. Let any later procedure, possibly randomized and computationally unbounded, output $A=a(E_m,Y,U)$, where $U$ is independent randomness. Because $A$ is computed only from $E_m$, $Y$, and $U$, these variables satisfy the conditional Markov relation $S\rightarrow E_m\rightarrow A$ given $Y$, so
\begin{equation}
I(S;A\mid Y)\le I(S;E_m\mid Y).
\label{eq:data-processing}
\end{equation}
This is the conditional data-processing inequality. Thus the population leakage objective upper-bounds the information available to any later rule for inferring the specified $S$. This includes, for example, membership, attribute, inversion, and architecture-inference attacks~\cite{shokri2017mia,yeom2018privacy,nasr2019whitebox,carlini2022lira,modelspy,deeptheft}. 

Fano's inequality then shows how the resulting evidence limits the success of any adversary attempting to infer $S$ from $(E_m,Y)$. For a binary $S$ that is balanced within each authorized class, it gives
\[
P_e\ge h_2^{-1}\!\left(1-I(S;E_m\mid Y)\right),
\]
where $P_e$ is the minimum error achievable by such an adversary and $h_2^{-1}$ is the lower branch of the inverse binary-entropy function. 



Only mechanisms that can be deployed are included:
\begin{equation}
\calM\subseteq
\{(c,\alpha,\theta):c\in\calC,\ \alpha\in\mathcal A_c,\
\theta\in\Theta_{c,\alpha}\}.
\label{eq:mechanism-set}
\end{equation}
The sets $\calC$, $\mathcal A_c$, and $\Theta_{c,\alpha}$ encode limits on cost and delay and the required security guarantees. MID can therefore select the sensor, change how data are collected, or compute a narrow release inside a protected device; it is not restricted to adding noise after a sensor has been fixed.

\paragraph{Finite-sample estimation.}

The population objective is defined with respect to the data-generating distribution, but in practice mechanisms must be compared using a finite development set. We therefore use either mutual-information estimators or finite-sample comparison scores appropriate to the candidate family. Mutual information measures how much each class-specific evidence distribution differs from the overall evidence distribution:
\begin{equation}
I(C;E_m)=\sum_c p(c)\KL\!\left(P(E_m\mid C=c)\,\|\,P(E_m)\right).
\label{eq:mi-mixture}
\end{equation}
Direct estimation is unreliable for high-dimensional traces and small datasets. Different types of evidence therefore require different estimators or finite-sample surrogates. For continuous traces, we use a regularized Gaussian separation score.

Suppose $E_m\mid C=c\approx\Normal(\mu_c,\Sigma_C)$ with shared covariance. We estimate class separation with the regularized Mahalanobis score
\begin{equation}
D_{a,b}(E_m)=(\mu_a-\mu_b)^\top(\Sigma_C+\gamma I)^{-1}(\mu_a-\mu_b).
\label{eq:pair-separation}
\end{equation}
Here $\gamma\ge0$ stabilizes covariance inversion. When $\gamma=0$, the quadratic form equals the sum of the two directed Kullback--Leibler (KL) divergences between shared-covariance Gaussians. For $\gamma>0$, it is a regularized separation proxy: regularization stabilizes the estimate without changing the released evidence.
For binary $Y$, we write
\begin{equation}
D_Y(E_m)=(\mu_1-\mu_0)^\top(\Sigma_Y+\gamma I)^{-1}(\mu_1-\mu_0),
\label{eq:dy}
\end{equation}
and use the weighted pairwise average for multiclass $Y$. Let $\mu_{s,y}$ denote the evidence mean for sensitive class $s$ at fixed authorized class $y$, and let $\Sigma_y$ be the corresponding pooled within-sensitive-class covariance. Conditional sensitive separability is computed within each authorized class:
\begin{equation}
D_S^{\mathrm{cond}}(E_m)=
\sum_y p(y)\!\sum_{s<s'}p(s\mid y)p(s'\mid y)
(\mu_{s,y}-\mu_{s',y})^\top
(\Sigma_y+\gamma I)^{-1}
(\mu_{s,y}-\mu_{s',y}).
\label{eq:ds}
\end{equation}
Under the balanced binary shared-covariance model, in the population limit with $\gamma=0$, the Bayes accuracy associated with authorized separation $D_Y$ is
\begin{equation}
a_Y=\Phi\!\left(\frac{\sqrt{D_Y}}{2}\right),
\qquad
D_Y\ge 4\bigl[\Phi^{-1}(a_Y)\bigr]^2.
\label{eq:gaussian-accuracy-threshold}
\end{equation}
where $\Phi$ is the standard-normal cumulative distribution function.
Thus a target accuracy can be translated into a separation threshold under this model; for example, $a_Y=0.80$ gives $D_Y\ge2.833$. This model-specific translation guides candidate comparison, and we also report held-out balanced accuracy (BA), the mean recall across classes. Because the score measures mean separation under shared covariance, we pair it with held-out prediction and attack evaluations that probe additional structure.

For reports with a small set of possible values, we compute plug-in mutual information from held-out frequencies and describe a zero value as zero measured leakage. Unless noted otherwise, authorized classes and sensitive classes within each authorized class receive equal prior weight. For rich evidence, information values derived from held-out prediction models are lower bounds on the information in the underlying trace.

\paragraph{Nested mechanism selection.}

Estimating every candidate on the same data used to select it would make both utility and leakage look better than they are. Algorithm~\ref{alg:mid} therefore splits the data twice. Inner splits fit and compare candidates; an outer group, unseen during selection, evaluates the chosen mechanism. Grouping is by physical run or matched configuration, so windows from one run and paired workloads never appear on both sides of a split.

\begin{algorithm}[H]
\caption{MID mechanism selection}
\label{alg:mid}
\begin{algorithmic}[1]
\State \textbf{Input:} episodes $\calD=\{(R_i,Y_i,S_i)\}$, specified candidates $\calM$, threshold $\tau_Y$, group labels
\For{each outer held-out group $G$}
  \For{each $m=(c,\alpha,\theta)\in\calM$}
    \State Fit $m$ on $\calD\setminus G$ using inner folds and generate inner out-of-fold releases
    \State Estimate $\widehat U_Y(m)$ and $\widehat L_S(m\mid Y)$ on those inner releases
  \EndFor
  \If{no candidate satisfies $\widehat U_Y(m)\ge\tau_Y$}
    \State \Return infeasible under the declared $\calM$ and $\tau_Y$
  \Else
    \State $\hat m_G\gets\arg\min_m\widehat L_S(m\mid Y)$ subject to $\widehat U_Y(m)\ge\tau_Y$
    \State Fit $\hat m_G$ on $\calD\setminus G$ and release evidence for $G$
  \EndIf
\EndFor
\State Combine the outer held-out results to evaluate the complete selection procedure
\State Re-run the same selection rule on all calibration groups and fit the final mechanism $m^\star$
\State Compile the interface for $m^\star$ as either an authenticated direct release or an authenticated private input with a ZKP relation for its transformation
\State \Return $m^\star$, its deployment specification, and outer-held-out measurements of the selection procedure
\end{algorithmic}
\end{algorithm}

The deployed monitor fixes $m^\star$ before observing a new execution and discloses only $E_{m^\star}$. The outer groups evaluate the complete process, including the choice among candidates; the final fit applies the same rule, declared in advance, to all development groups. Mutual information supplies the common objective; Gaussian scores and prediction models support finite-sample selection, while attack evaluations provide separate empirical checks.

\section{Concrete MID Mechanisms And Deployment}
\label{sec:deployment}

The information-theoretic objective determines what a selected evidence interface must achieve; this section turns that choice into concrete mechanisms and deployable releases. We first describe mechanism families that act during or after collection, then show how authenticated sources and ZKPs bind the selected release to the execution.

\subsection{Concrete Mechanisms}

We instantiate $\calM$ with reusable mechanism families that act either during collection or after it. Section~\ref{sec:experiments} specifies the channel $c$, collection policy $\alpha$, and release rule $\theta$ selected in each task.

\paragraph{Collection-time mechanisms.}
Channel restriction selects which physical signal, telemetry field, counter, or static hardware attribute the evidence source acquires. Acquisition minimization controls the sampling rate, observation window, sensor placement, or summary computed inside the sensor. These choices directly construct $X_{c,\alpha}$; when the result meets the utility threshold, the disclosed evidence is $E_m=X_{c,\alpha}$ and no richer deployment trace is created.

\paragraph{Hard-decision and quantized reports.}
When $X_{c,\alpha}$ remains more informative than the authorized claim requires, a post-collection rule can disclose a coarser result. Given an authorized score $h(X_{c,\alpha})$, a hard-decision mechanism releases
\begin{equation}
Z_{\mathrm{hard}}=\mathbb{I}\!\left\{h(X_{c,\alpha})\ge t\right\}.
\label{eq:hard-release}
\end{equation}
Here $t$ is the hard-decision threshold.
A quantized randomized mechanism instead releases
\begin{equation}
Z_{q}=q\,\mathrm{round}\!\left(
\frac{h(X_{c,\alpha})+\eta}{q}
\right),
\qquad q>0,\quad \eta\sim\Normal(0,\sigma^2),
\label{eq:quantized-release}
\end{equation}
which includes deterministic quantization when $\sigma=0$. These transformations can be combined with a selected collection-time mechanism and computed inside a protected device after collection.

\paragraph{Privacy-aware linear projections.}

Hard decisions and fixed quantizers may not yield a feasible point under Eq.~\eqref{eq:operational}. A more flexible post-collection family is the $k$-dimensional noisy linear release
\begin{equation}
Z=A_\beta X_{c,\alpha}+\sigma\xi,
\qquad \xi\sim\Normal(0,I_k).
\label{eq:linear-release}
\end{equation}
For binary $Y$, let $\mu_y=\mathbb E[X_{c,\alpha}\mid Y=y]$, let $\Delta_Y=\mu_1-\mu_0$, and let $\Sigma_Y$ be the pooled within-$Y$ covariance. With $\mu_{s,y}=\mathbb E[X_{c,\alpha}\mid S=s,Y=y]$, define conditional sensitive scatter
\begin{equation}
B_{S\mid Y}=\sum_y p(y)\sum_s p(s\mid y)
(\mu_{s,y}-\mu_y)(\mu_{s,y}-\mu_y)^\top.
\label{eq:sensitive-scatter}
\end{equation}
For a covariance-regularization parameter $\lambda>0$, the rows of $A_\beta$ are the leading generalized eigenvectors $v$ satisfying
\begin{equation}
\Delta_Y\Delta_Y^\top v
=\lambda_v(\Sigma_Y+\beta B_{S\mid Y}+\lambda I)v.
\label{eq:generalized-eigen}
\end{equation}
Here $\lambda_v$ is the generalized eigenvalue associated with $v$.
For $k=1$, the release direction is proportional to
\begin{equation}
w_\beta=(\Sigma_Y+\beta B_{S\mid Y}+\lambda I)^{-1}\Delta_Y.
\label{eq:linear-direction}
\end{equation}
$\beta$ penalizes directions that distinguish $S$ within fixed $Y$; $\sigma$ adds release noise.

\subsection{Verifiable Deployment With ZKPs}
\label{sec:integrity}

The mechanisms above determine what should cross the trust boundary; deployment must also establish that the evidence came from the claimed execution and that any post-collection transformation was applied correctly. Both paths require an authenticated evidence source, while the post-collection path additionally uses a ZKP to certify the selected transformation without revealing the private witness beyond what follows from the public instance. This computation-integrity guarantee holds for a computationally bounded verifier under the ZKP system's cryptographic and setup assumptions. Groth16~\cite{groth2016} additionally requires a circuit-specific structured reference string; a deployment must generate it through an appropriate trusted or multi-party setup and bind the proving and verification keys to the audited circuit.
Let $m^\star=(c^\star,\alpha^\star,\theta^\star)$. The evidence source commits to $X_{c^\star,\alpha^\star}$ and links that commitment to an execution identifier $\mathrm{id}_W$ and the selected collection configuration. The values given to the verifier are called the public instance:
\[
x_{\mathrm{pub}}=(\mathrm{id}_W,C_X,Z_{\theta^\star},m^\star,\sigma_{\mathrm{src}}),
\]
where $\sigma_{\mathrm{src}}$ is the source-authentication signature or tag. Here \emph{public} means visible to the verifier outside the ZKP; it does not require unrestricted publication.
The values kept hidden inside the ZKP, called the private witness, are $w=(X_{c^\star,\alpha^\star},\xi,r)$, where $r$ opens the commitment. A deployment accepts only if
\begin{equation}
\begin{aligned}
\exists X_{c^\star,\alpha^\star},\xi,r:\quad
&C_X=\mathrm{Com}(X_{c^\star,\alpha^\star};r)\\
&\land\ \mathrm{VerifySrc}\!\left(
  \sigma_{\mathrm{src}};
  \mathrm{id}_W,C_X,c^\star,\alpha^\star
\right)=1\\
&\land\ Z_{\theta^\star}=g_{\theta^\star}(X_{c^\star,\alpha^\star},\xi).
\end{aligned}
\label{eq:zk-statement}
\end{equation}
The source-authentication check may be evaluated inside the circuit or separately. In either case, acceptance also requires
\begin{equation}
\mathrm{ZK.Verify}(vk,x_{\mathrm{pub}},\pi)=1.
\label{eq:zk-verify}
\end{equation}
Here $vk$ is the verification key and $\pi$ is the ZKP. The monitor sends $(\mathrm{id}_W,C_X,Z_{\theta^\star},m^\star,\sigma_{\mathrm{src}},\pi)$ to the verifier. Computational zero knowledge protects $X_{c^\star,\alpha^\star}$, $\xi$, and $r$ beyond what follows from the public instance. The source signature establishes measurement origin and integrity, while the commitment and ZKP bind the released value to the private measurement and selected computation. All verifier-visible fields are included in $E_m$ for disclosure accounting.

For a linear scalar release, a fixed-point circuit checks
\begin{equation}
\widetilde Z=
\sum_{j=1}^{d}\widetilde w_j\widetilde X_j
+\widetilde\sigma\widetilde\xi.
\label{eq:zk-fixed}
\end{equation}
Tildes denote fixed-point integer encodings, and $d$ is the measurement-vector dimension. The mechanism parameters and $\widetilde Z$ are public; $\widetilde X$ and $\widetilde\xi$ are private. A deployment circuit also enforces signed fixed-point encodings, scaling, rounding, and range bounds so that field equality matches the intended integer arithmetic. Randomized releases additionally require verifiable sampling; for example, the prover can commit to a hidden seed before a verifier challenge and derive finite-precision noise inside the circuit. The public commitment and challenge are included in $E_m$, while the seed and noise remain private so that the verifier cannot subtract the noise from the release.
Section~\ref{sec:deeptheft} demonstrates the arithmetic in Eq.~\eqref{eq:zk-fixed} in Circom~\cite{circom} and uses snarkjs~\cite{snarkjs} to generate and verify a Groth16 zk-SNARK~\cite{groth2016} over the BN254 pairing-friendly curve, which snarkjs calls \texttt{bn128}.

\section{Empirical Evaluation}
\label{sec:experiments}
Having defined MID's objective, mechanisms, and deployment paths, we now ask whether it works in practice: Can MID select deployable evidence that preserves the information needed to verify a policy-relevant claim while suppressing collateral leakage?

\subsection{Experimental Setup}

To test MID across different policy questions and evidence sources, we evaluate six tasks on four public physical-measurement datasets, organized by the question answered by $Y$. Table~\ref{tab:roadmap} gives the policy motivation and measurement channel. Each case defines $S$, compares deployable mechanisms, and reports either a utility-constrained selection or the measured frontier from which a policy would select. The experiments treat the dataset records as authentic and focus on disclosure; Section~\ref{sec:integrity} specifies deployment checks for origin, integrity, coverage, freshness, and execution binding. Throughout the evaluation, BA denotes balanced accuracy, the mean recall across classes. A score is \emph{out-of-fold} when it is produced by a model that was not trained on the physical run being scored.

\begin{table}[!tbp]
\centering
\caption{Verification questions evaluated in this paper.}
\label{tab:roadmap}
{\footnotesize
\setlength{\tabcolsep}{3pt}
\renewcommand{\arraystretch}{1.05}
\begin{tabularx}{\textwidth}{@{}>{\raggedright\arraybackslash}p{0.19\textwidth}>{\raggedright\arraybackslash}p{0.28\textwidth}>{\raggedright\arraybackslash}p{0.30\textwidth}>{\raggedright\arraybackslash}X@{}}
\toprule
Question & Why it matters & Authorized target $Y$ & Measurement signal\\
\midrule
\textbf{Execution Type}\\Is the allowed kind of computation running? & Inference-only proposals require evidence that distinguishes training from inference; policy and safety work also distinguishes consequential post-training methods~\cite{shavit2023chinchilla,rahman2026hidden,ai2040verification}. & Training vs. inference (Sec.~\ref{sec:training-inference}); reinforcement learning (RL) vs. non-RL training (Sec.~\ref{sec:rl}) & Facility electrical power; central processing unit (CPU) and graphics processing unit (GPU) utilization telemetry\\
\textbf{Hardware Identity}\\Is the approved accelerator being used? & Chip controls, operating licenses, and end-use verification require trustworthy claims about advanced-computing hardware~\cite{aarne2024chips,ogara2025hem,ansari2026taxonomy,bis2025chips}. & H100 vs. B200 (Sec.~\ref{sec:hardware}) & Accelerator memory, power, utilization, and temperature telemetry\\
\textbf{Compute Scale}\\Is resource use within the allowed class? & Compute-governance and international-verification proposals require evidence about the amount and organization of accelerator use~\cite{sastry2024compute,shavit2023chinchilla,harack2025verification,scher2025mechanisms,euai2025gpai}. & Low vs. high execution tier (Sec.~\ref{sec:deeptheft}); number of compute nodes / number of GPUs used (Sec.~\ref{sec:nodes}) & Processor-package and memory power; facility electrical power\\
\textbf{Model Identity}\\Is the approved model type running? & Inference-verification and property-attestation systems require evidence that a claimed model or authorized variant actually executed~\cite{toploc2025,karvonen2025difr,cankaya2026bitexact,chantasantitam2026palm}. & Convolutional neural network (CNN) vs. Transformer (Sec.~\ref{sec:modelspy}) & GPU electromagnetic trace\\
\bottomrule
\end{tabularx}}
\end{table}

\subsection{Execution Type}

We begin with execution-type verification, which asks what computation is occurring without disclosing the finer workload. We study training versus inference and, within training, reinforcement learning versus non-RL procedures.

\paragraph{Training versus inference.}
\label{sec:training-inference}

Training detection is a direct prerequisite for policies that distinguish restricted training from permitted inference. Rahman and Tajdari evaluate this distinction using GPU telemetry under explicit evasion strategies~\cite{rahman2026hidden}. We study the same authorized question through a different physical channel. Our processed corpus, constructed from the National Laboratory of the Rockies (NLR) dataset, contains Llama-2 70B low-rank adaptation (LoRA) and Stable Diffusion training runs together with Llama-3.1 70B inference runs from the InstructCoder and MLPerf prompt collections~\cite{nlr2026power}. We authorize
\[
Y\in\{\text{training},\text{inference}\},
\]
and treat the finer workload identity as sensitive: Llama-2 versus Stable Diffusion within training and InstructCoder versus MLPerf within inference.

The raw baseline contains 27 facility-power summaries. Candidate reports vary the feature family, quantization width, and Gaussian noise. The selected feature family contains six dimensionless shape statistics---coefficient of variation, peak-to-mean ratio, autocorrelation at lags 1, 5, and 10, and spectral entropy---and excludes absolute power, duration, energy, sample count, and sampling interval. The report uses Eq.~\eqref{eq:quantized-release} with the out-of-fold binary classifier $h(X)=\widehat Y_X$, $q=0.05$, and $\sigma=0.15$. We average the reported metrics over the mechanism's randomness rather than report one noise draw.

\begin{figure}[!tbp]
\centering
\includegraphics[width=\textwidth]{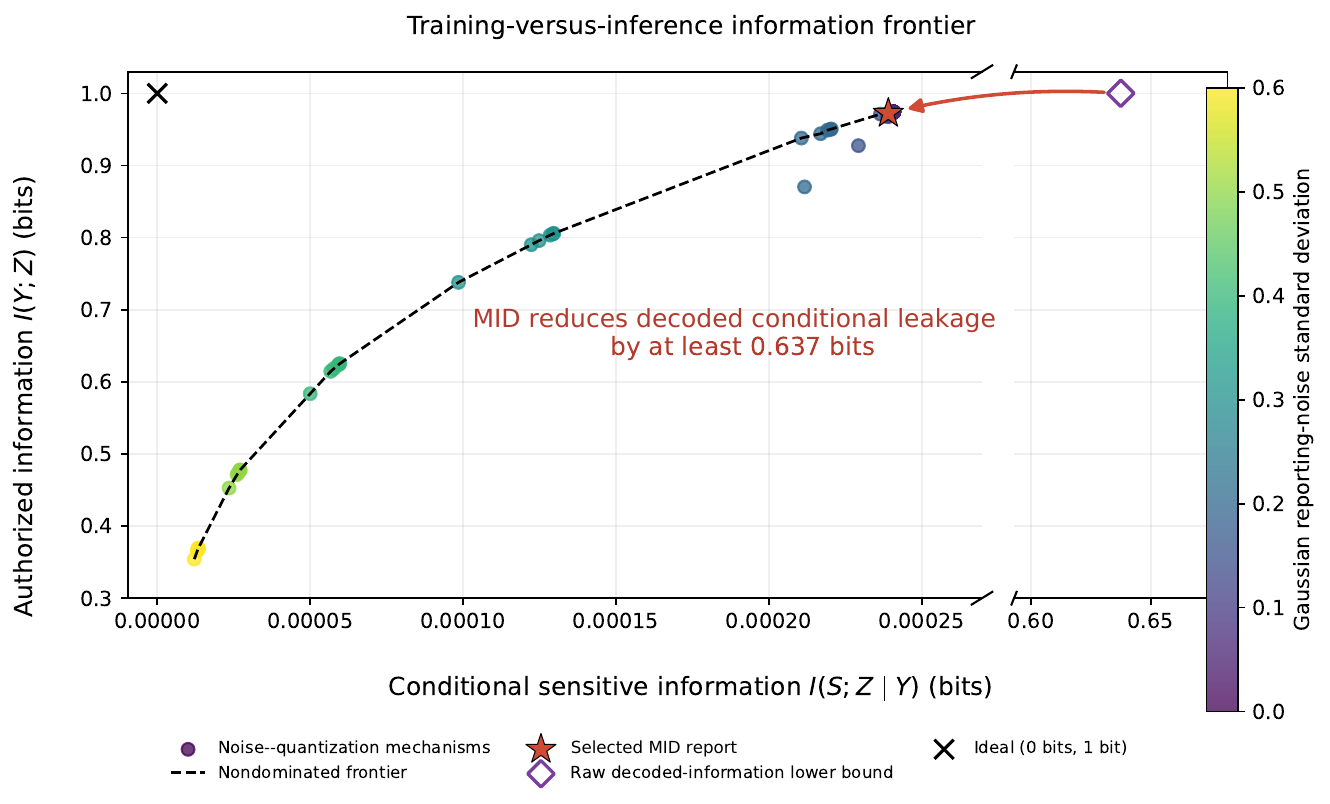}
\caption{Training-versus-inference information frontier. The selected report retains $0.9723$ bits of authorized information while reducing measured conditional leakage from a raw-evidence lower bound of $0.6375$ bits to $0.000239$ bits. Its position near the upper-left ideal shows that MID removes workload information without making the evidence uninformative for verification.}
\label{fig:training-frontier}
\end{figure}
\begin{figure}[!tbp]
\centering
\includegraphics[width=\textwidth]{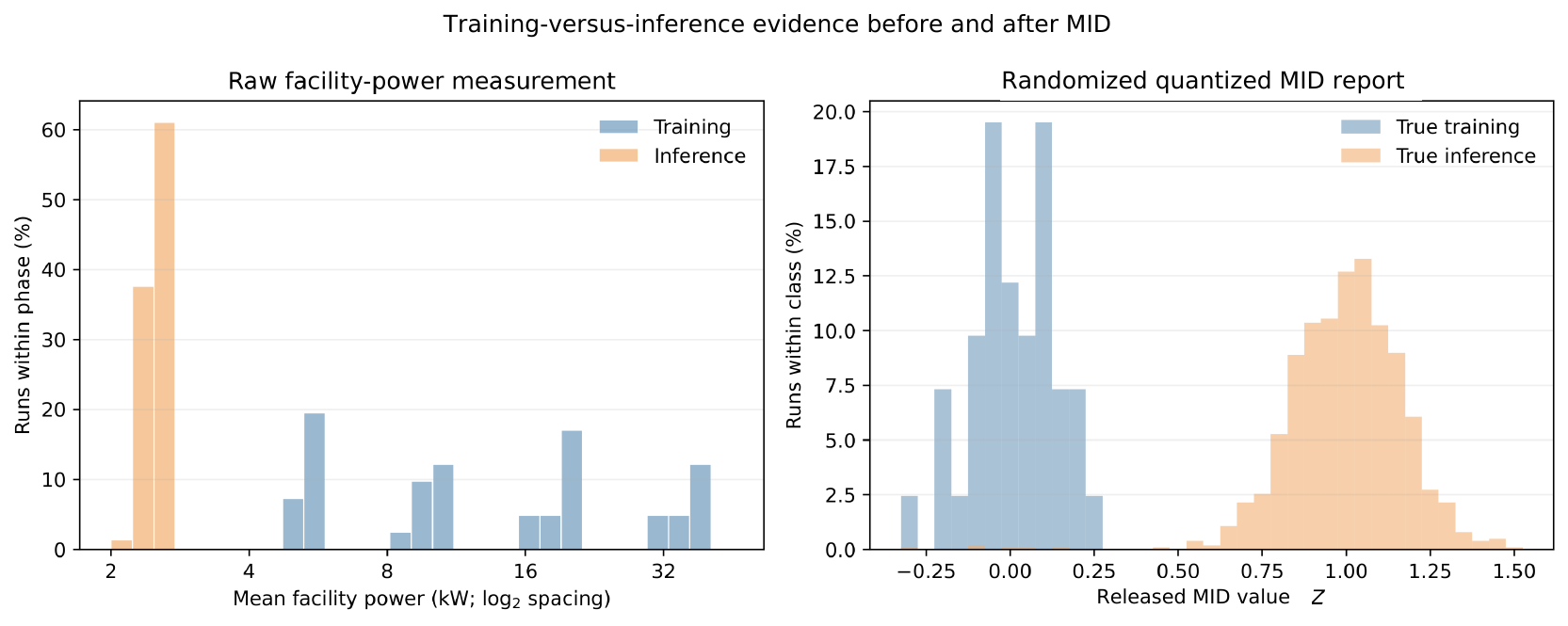}
\caption{Training-versus-inference evidence before and after MID. The selected randomized report retains separated, multi-bin distributions for the authorized classes.}
\label{fig:training-y-detail}
\end{figure}
\begin{figure}[!tbp]
\centering
\includegraphics[width=0.9\textwidth]{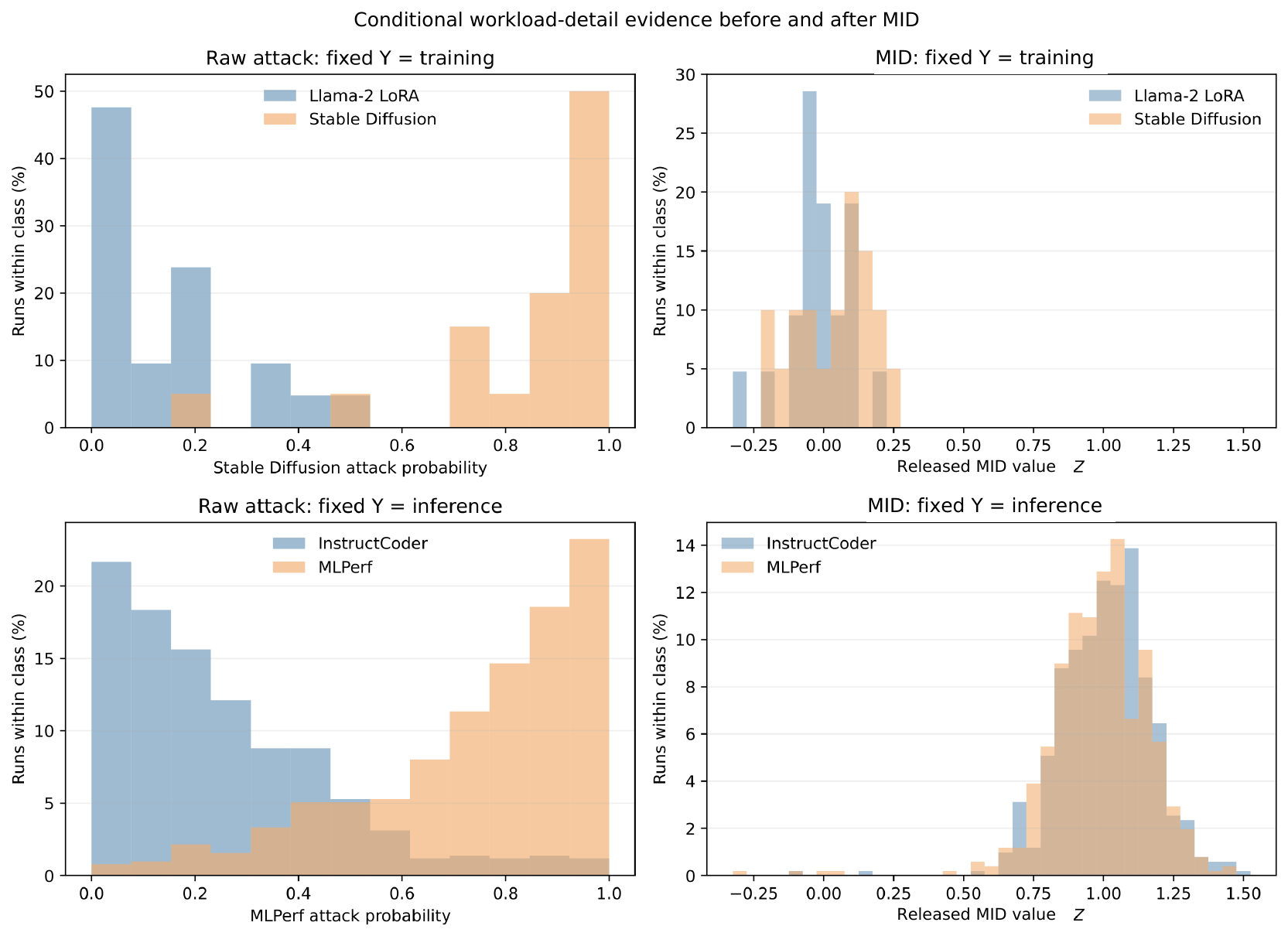}
\caption{Conditional workload evidence. The top row fixes training and compares Llama-2 LoRA with Stable Diffusion; the bottom row fixes inference and compares InstructCoder with MLPerf. Raw trace features distinguish each sensitive pair, whereas the selected report makes the corresponding distributions nearly identical.}
\label{fig:training-s-detail}
\end{figure}

In grouped out-of-fold evaluation, the raw features distinguish training from inference perfectly but also support sensitive-workload inference with mean balanced accuracy $0.9182$. The selected MID report retains $Y$ balanced accuracy $0.9966$ and $I(Y;Z)=0.9723$ bits, while reducing sensitive inference to $0.5010$ and measured conditional leakage to $0.000239$ bits. Table~\ref{tab:training-results} therefore shows that the workload distinctions contained in raw facility-power evidence are not necessary for training-versus-inference verification. Figure~\ref{fig:training-frontier} shows how MID places the selected report close to the ideal of high authorized information and zero collateral leakage. Figure~\ref{fig:training-y-detail} shows that the authorized classes remain separated after MID, while Figure~\ref{fig:training-s-detail} shows that the protected workload distributions become nearly indistinguishable within both training and inference, hence minimal leakage.

\begin{table}[!tbp]
\centering
\caption{Training-versus-inference results. MID retains $0.9966$ authorized balanced accuracy while reducing mean sensitive-workload balanced accuracy from $0.9182$ to $0.5010$ and measured conditional leakage to $0.000239$ bits.}
\label{tab:training-results}
\resizebox{\textwidth}{!}{%
\begin{tabular}{lcccccc}
\toprule
Evidence & $Y$ BA & $S\mid Y{=}\mathrm{train}$ BA & $S\mid Y{=}\mathrm{infer}$ BA & Mean $S\mid Y$ BA & $I(Y;E)$ & $I(S;E\mid Y)$\\
\midrule
Raw trace features & 1.0000 & 0.9750 & 0.8613 & 0.9182 & 1.0000 & 0.6375\\
Scale-invariant MID report & 0.9966 & 0.5000 & 0.5020 & 0.5010 & 0.9723 & 0.000239\\
\bottomrule
\end{tabular}}
\end{table}

\paragraph{Reinforcement learning versus non-RL training.}
\label{sec:rl}

Reinforcement-learning post-training can materially change model behavior and capabilities, so whether RL was used is itself worth verifying~\cite{deepseekr1,iaisr2026}. We ask whether telemetry can support that claim without disclosing the exact non-RL workload.

After removing byte-identical copies, the single-machine Elsayed dataset contains physical sessions spanning RL, forecasting, image captioning, image classification, and text generation~\cite{elsayed2026traces}. We authorize $Y\in\{\text{RL},\text{non-RL}\}$ and, at fixed $Y=\text{non-RL}$, protect the four-way exact workload. After excluding startup and shutdown edges, each session is divided into non-overlapping 300-sample windows. All windows from a session remain in the same one of four held-out folds. Models are fitted on windows, but held-out probabilities are averaged within each session before computing any metric or information value.

The rich baseline contains 144 summaries of GPU and CPU power, utilization, memory, temperature, and voltage. Candidate channels include individual and combined power or utilization, GPU and system memory, temperature, voltage, and scale-independent temporal shape. The selected channel contains three fields: GPU core load, activity through NVIDIA's CUDA computing interface, and mean Direct3D (D3D) utilization. If $\widehat p_i$ is the held-out session-level RL probability, the release is
\begin{equation}
Z_i=\mathbb{I}\{\widehat p_i\ge 1/2\},
\qquad 0\equiv\text{non-RL},\quad 1\equiv\text{RL}.
\label{eq:rl-report}
\end{equation}
A classifier using all 144 telemetry summaries recovers both $Y$ and the four-way non-RL workload perfectly, giving one bit of authorized information and two bits of sensitive information under equal priors. The one-bit release is correct for every held-out session and constant across non-RL workloads: $I(Y;Z)=1$ bit, sensitive balanced accuracy is $0.25$, and the held-out plug-in estimate is $I(S;Z\mid Y=\text{non-RL})=0$.

\begin{figure}[!tbp]
\centering
\includegraphics[width=\textwidth]{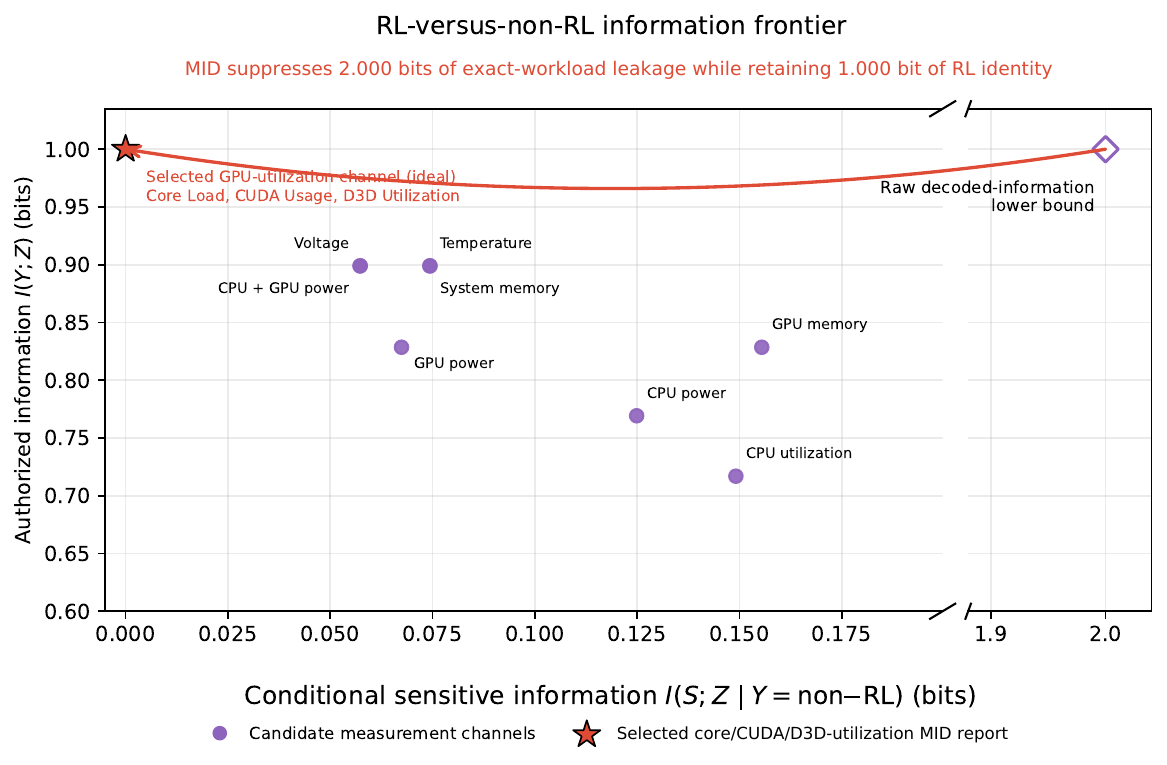}
\caption{RL-versus-non-RL information frontier across telemetry-channel mechanisms. Every colored candidate is a one-bit RL-status report, yet several have nonzero exact-workload leakage. On the held-out sessions, the selected GPU-utilization report retains one bit of authorized information with zero measured exact-workload information at fixed non-RL.}
\label{fig:rl-frontier}
\end{figure}

\begin{table}[!tbp]
\centering
\caption{RL-versus-non-RL results. The sensitive target is the four-way exact workload, evaluated only for non-RL sessions.}
\label{tab:rl-results}
\begin{tabular}{lcccc}
\toprule
Evidence & $Y$ BA & $I(Y;E)$ & $S$ BA & $I(S;E\mid Y{=}\mathrm{non\text{-}RL})$\\
\midrule
Raw telemetry & 1.0000 & 1.0000 & 1.0000 & 2.0000\\
Selected GPU-utilization report & 1.0000 & 1.0000 & 0.2500 & 0\\
\bottomrule
\end{tabular}
\end{table}
\FloatBarrier
Each colored candidate in Figure~\ref{fig:rl-frontier} is a one-bit hard report constructed from a different telemetry channel; the diamond is the rich-telemetry decoder lower bound. Several one-bit candidates still reveal exact-workload information because their errors vary across workloads at fixed non-RL. A small output alphabet therefore does not by itself control disclosure. The frontier compares measurement channels, not only sample rates or noise levels within one predetermined channel.

\begin{figure}[!tbp]
\centering
\includegraphics[width=\textwidth]{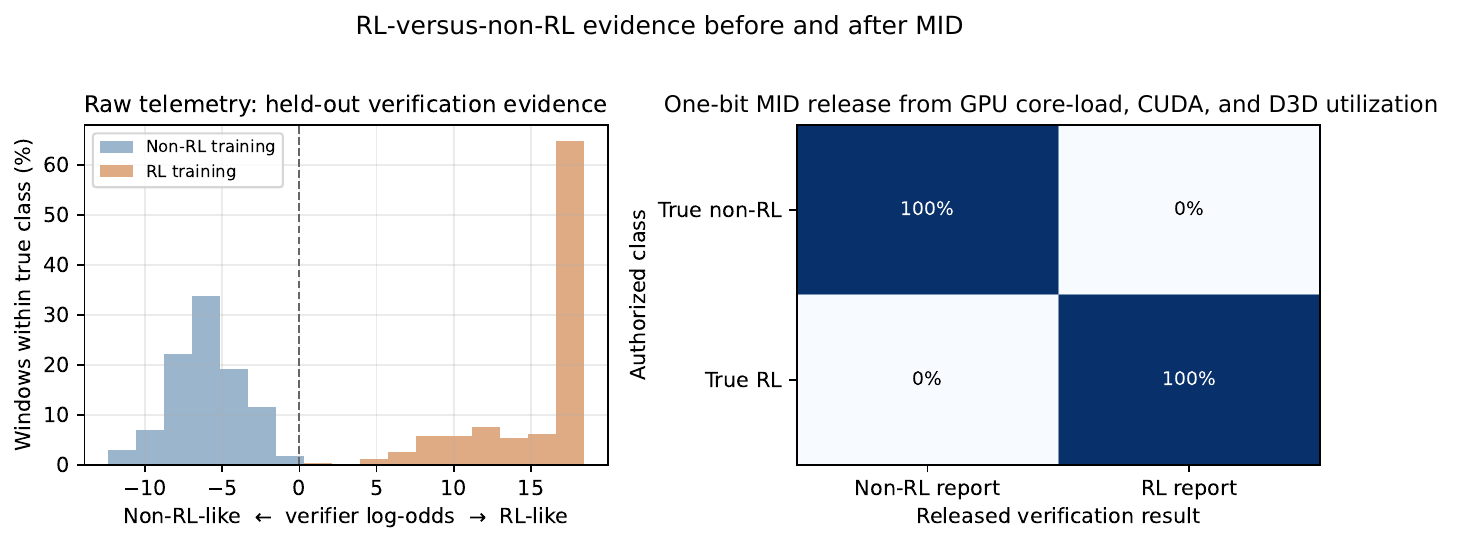}
\caption{RL-versus-non-RL held-out verification evidence and the selected one-bit report. The left panel shows a classifier score (log-odds) computed from rich telemetry; the right panel shows the released RL/non-RL bit.}
\label{fig:rl-y-detail}
\end{figure}

\begin{figure}[!tbp]
\centering
\includegraphics[width=\textwidth]{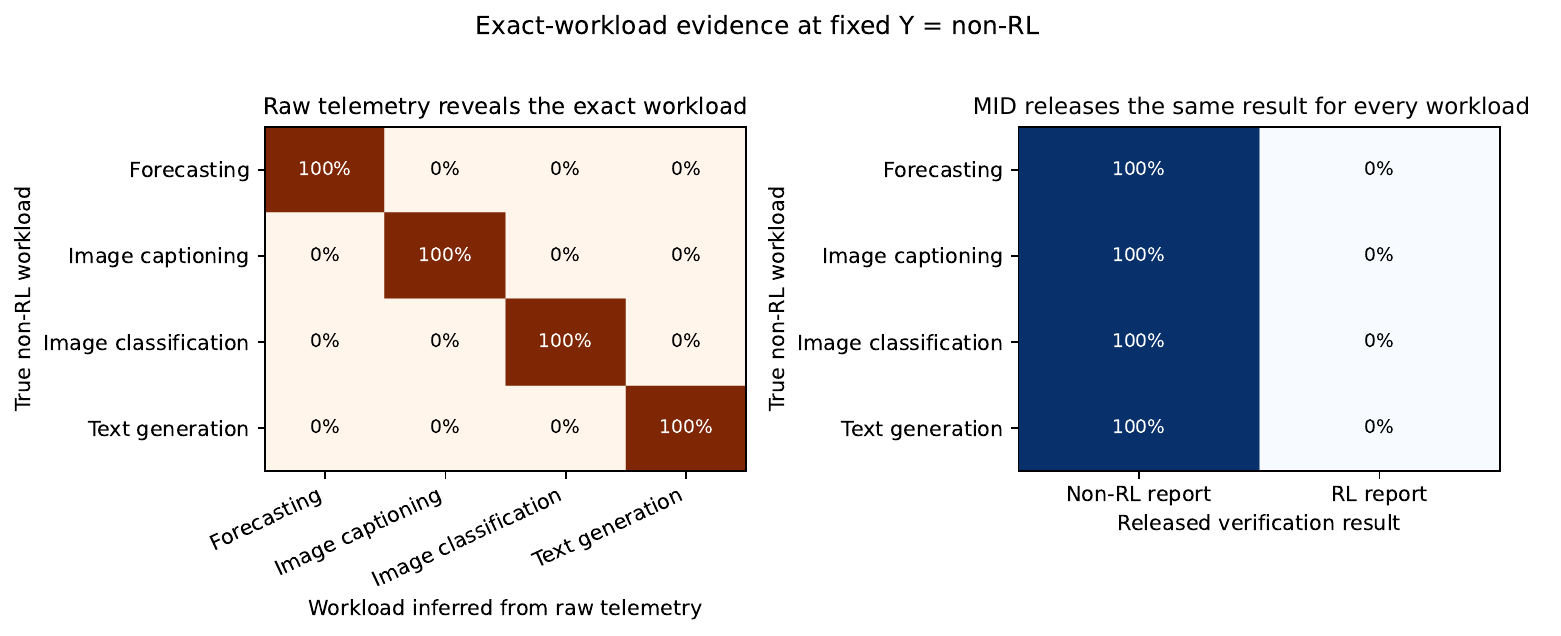}
\caption{Exact-workload evidence at fixed non-RL. Raw telemetry identifies forecasting, image captioning, image classification, and text generation; MID releases the same non-RL result for all four workloads.}
\label{fig:rl-s-detail}
\end{figure}

Figures~\ref{fig:rl-y-detail} and~\ref{fig:rl-s-detail} explain why the MID selected report reaches the empirical ideal. It is perfectly correct for RL versus non-RL (Figure~\ref{fig:rl-y-detail}) and returns the \textit{same} non-RL answer for forecasting, image captioning, image classification, and text generation (Figure~\ref{fig:rl-s-detail}). Other candidate one-bit reports make workload-dependent errors and therefore retain exact-workload information. The result is not that one-bit reports are inherently private; it is that MID identifies which one-bit report preserves the authorized answer without encoding the protected workload through its behavior.

\subsection{Hardware Identity}
\label{sec:hardware}

We next apply MID to hardware identity: whether an approved accelerator executed the workload without revealing the workload class. After removing byte-identical copies, we use diffusion and large language model (LLM) configurations from Elsayed et al. that were measured on both eight-GPU H100 and B200 nodes~\cite{elsayed2026traces}. RTX~3060 traces lack matched eight-GPU configurations and are excluded. Each H100/B200 pair shares one of five held-out folds. The authorized target is chip identity; the sensitive target is diffusion versus LLM within each chip.

The raw baseline contains 183 summaries of power, utilization, memory, and temperature. Candidate mechanisms use full telemetry, dynamic telemetry without static capacity, GPU power, utilization, memory activity, temperature, scale-independent temporal shape, or reported memory capacity. MID selects total reported GPU memory capacity
\[
V_i=\operatorname{mean}_t\sum_{g=1}^{8}\mathrm{memTotal}_{i,g}(t).
\]
Inside fold $f$, the threshold is fitted using training sessions only:
\[
b_f=\frac{1}{2}\left(
\max_{i\in\mathrm{train}_f:Y_i=\mathrm{H100}}V_i+
\min_{i\in\mathrm{train}_f:Y_i=\mathrm{B200}}V_i
\right).
\]
The released evidence is
\begin{equation}
Z_i=
\begin{cases}
0, & V_i\le b_f\quad(\mathrm{H100}),\\
1, & V_i>b_f\quad(\mathrm{B200}).
\end{cases}
\label{eq:hardware-report}
\end{equation}
When each matched group is scored by a model trained on the other groups, the telemetry verifies the chip perfectly but identifies diffusion versus LLM with balanced accuracy $0.8611$ on H100 and $0.8889$ on B200. The one-bit report preserves perfect chip verification and is constant across workloads within each chip. This analysis assumes that reported capacity is authentic; deployment requires the authenticated source specified in Section~\ref{sec:problem}.

\begin{table}[!tbp]
\centering
\caption{H100-versus-B200 results. Raw telemetry verifies chip identity perfectly but also distinguishes diffusion from LLM workloads with $0.8750$ BA. The MID-selected memory-capacity report preserves perfect chip verification while reducing workload inference to chance and measured conditional leakage from $0.5016$ bits to zero.}
\label{tab:hardware-results}
\resizebox{\textwidth}{!}{%
\begin{tabular}{lcccccc}
\toprule
Evidence & $Y$ BA & $S\mid Y{=}\mathrm{H100}$ BA & $S\mid Y{=}\mathrm{B200}$ BA & Mean $S\mid Y$ BA & $I(Y;E)$ & $I(S;E\mid Y)$\\
\midrule
Raw telemetry & 1.0000 & 0.8611 & 0.8889 & 0.8750 & 1.0000 & 0.5016\\
Memory-capacity MID report & 1.0000 & 0.5000 & 0.5000 & 0.5000 & 1.0000 & 0\\
\bottomrule
\end{tabular}}
\end{table}

\begin{figure}[!tbp]
\centering
\includegraphics[width=0.82\textwidth]{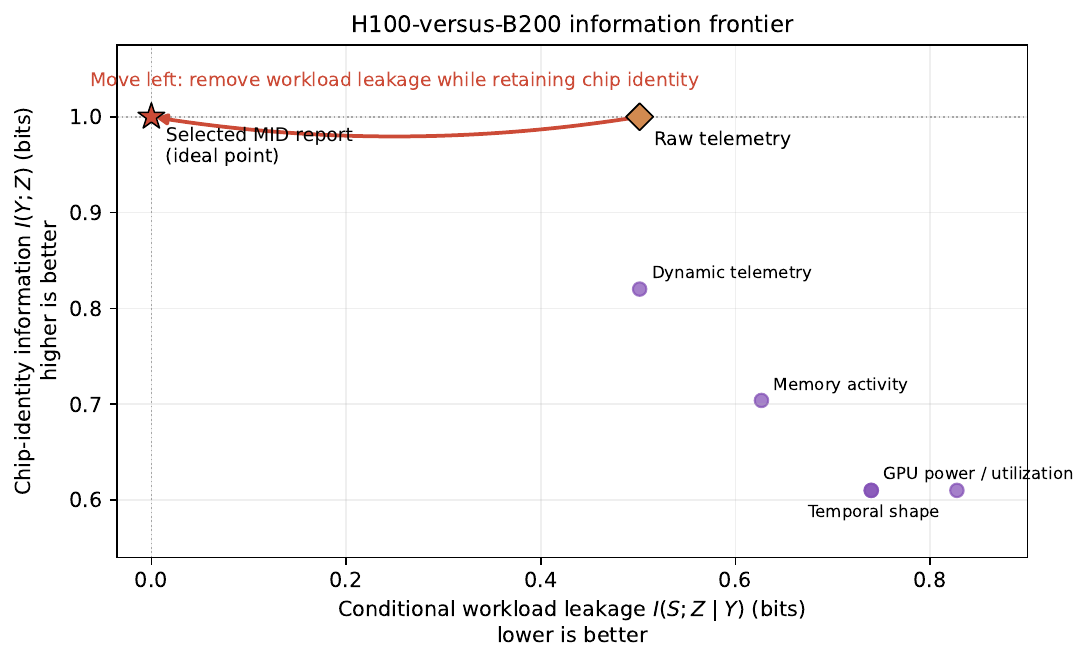}
\caption{H100-versus-B200 information frontier. Each candidate uses a different telemetry channel or channel family. The experiment shows that MID can choose what to measure, not only how to alter a fixed measurement.}
\label{fig:hardware-frontier}
\end{figure}

Table~\ref{tab:hardware-results} and Figure~\ref{fig:hardware-frontier} show that the dynamic behavior recorded in rich telemetry is unnecessary for this authorized claim. Raw telemetry and authenticated memory capacity both retain one full bit of chip-identity information, but only the dynamic telemetry exposes substantial workload information. By searching across evidence channels, MID finds that a static capacity report reaches the empirical ideal, whereas several dynamic channels lose chip utility while retaining more workload leakage.

\begin{figure}[!tbp]
\centering
\includegraphics[width=\textwidth]{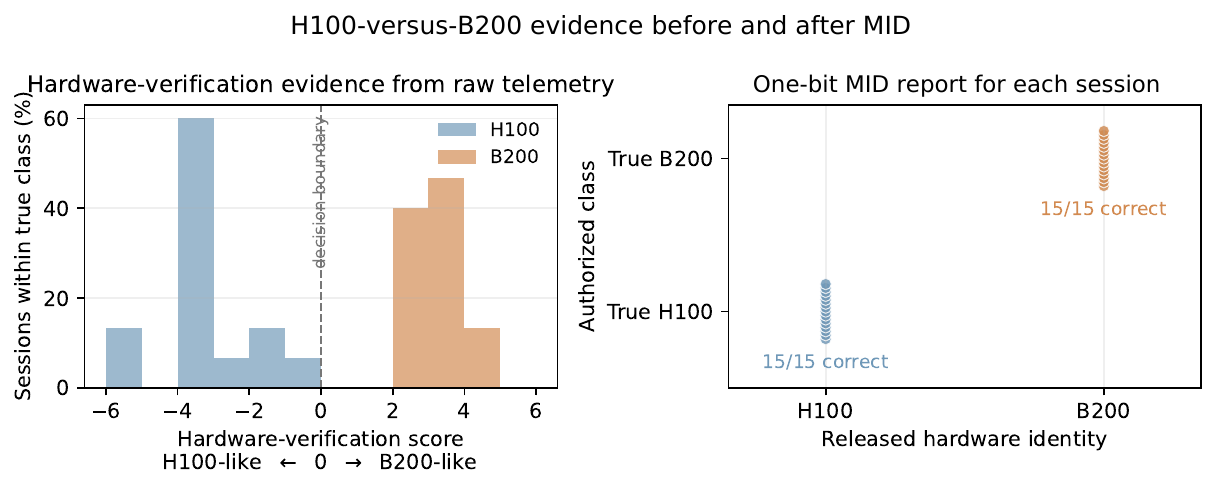}
\caption{H100-versus-B200 verification before and after MID. The left panel visualizes the multivariate raw telemetry with a held-out log-odds score; the right panel shows the actual one-bit hardware report. Both rich telemetry and the selected one-bit capacity report identify the chip perfectly, showing that the richer dynamic trace is unnecessary for the authorized hardware claim.}
\label{fig:hardware-y-detail}
\end{figure}

\begin{figure}[!tbp]
\centering
\includegraphics[width=\textwidth]{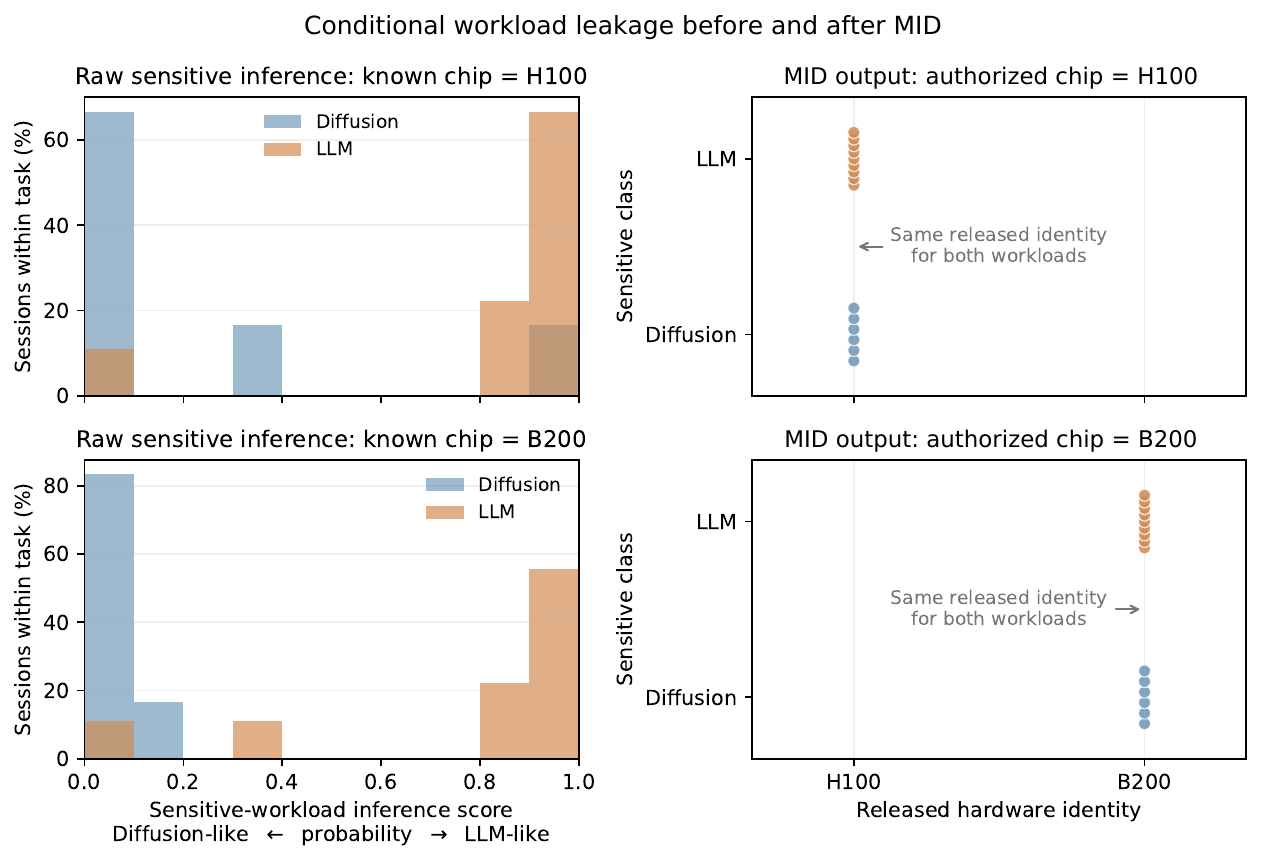}
\caption{Conditional diffusion-versus-LLM evidence within each authorized chip. Rich telemetry distinguishes the workloads, while both workloads induce the same MID released chip result, hence zero leakage.}
\label{fig:hardware-s-detail}
\end{figure}

Figures~\ref{fig:hardware-y-detail} and~\ref{fig:hardware-s-detail} show why this channel choice works. The capacity report remains perfectly separated by H100 versus B200, but within either chip it is identical for diffusion and LLM workloads. MID therefore removes the workload distinction without weakening the hardware-identity claim.

\subsection{Compute Scale}

We then turn to compute-scale verification, where the policy may target a coarse resource class or an exact declared allocation. We evaluate both using processor-level and facility-level power.

\paragraph{Coarse execution tier using DeepTheft traces.}
\label{sec:deeptheft}

Compute-monitoring proposals seek to verify resource limits without exposing models, data, or hyperparameters~\cite{shavit2023chinchilla,sastry2024compute}. The dataset accompanying DeepTheft contains variable-length processor-package and dynamic random-access memory (DRAM) power traces collected through Intel's Running Average Power Limit (RAPL) interface during deep neural network executions~\cite{deeptheft}. We define $Y$ from trace duration: the bottom 30\% is low tier, the top 30\% high tier, and the middle 40\% is excluded. This laboratory target tests disclosure control for a coarse duration class; it is not a proxy for statutory training-compute thresholds expressed as a number of floating-point operations (FLOP) or for frontier-training scale.

We treat model family and architecture as separate sensitive targets and design a release for each. The first sensitive target contains six model families,
\[
S_{\mathrm{family}}\in\{\texttt{custom\_net},\texttt{custom\_net\_bn},
\texttt{resnet\_basicblock},\texttt{resnet\_bottleneck},
\texttt{vgg},\texttt{vgg\_bn}\}.
\]
The second is the architecture tuple
\[
S_{\mathrm{arch}}=(n_{\mathrm{layers}},n_{\mathrm{conv}},n_{\mathrm{pool}},
n_{\mathrm{linear}},\mathrm{max\ output\ channels}),
\]
which yields five classes after retaining only architecture categories represented in both compute tiers. Architecture metadata supplies labels only; no architecture field is an attack input or released feature.

\begin{figure}[!tbp]
\centering
\includegraphics[width=0.9\textwidth]{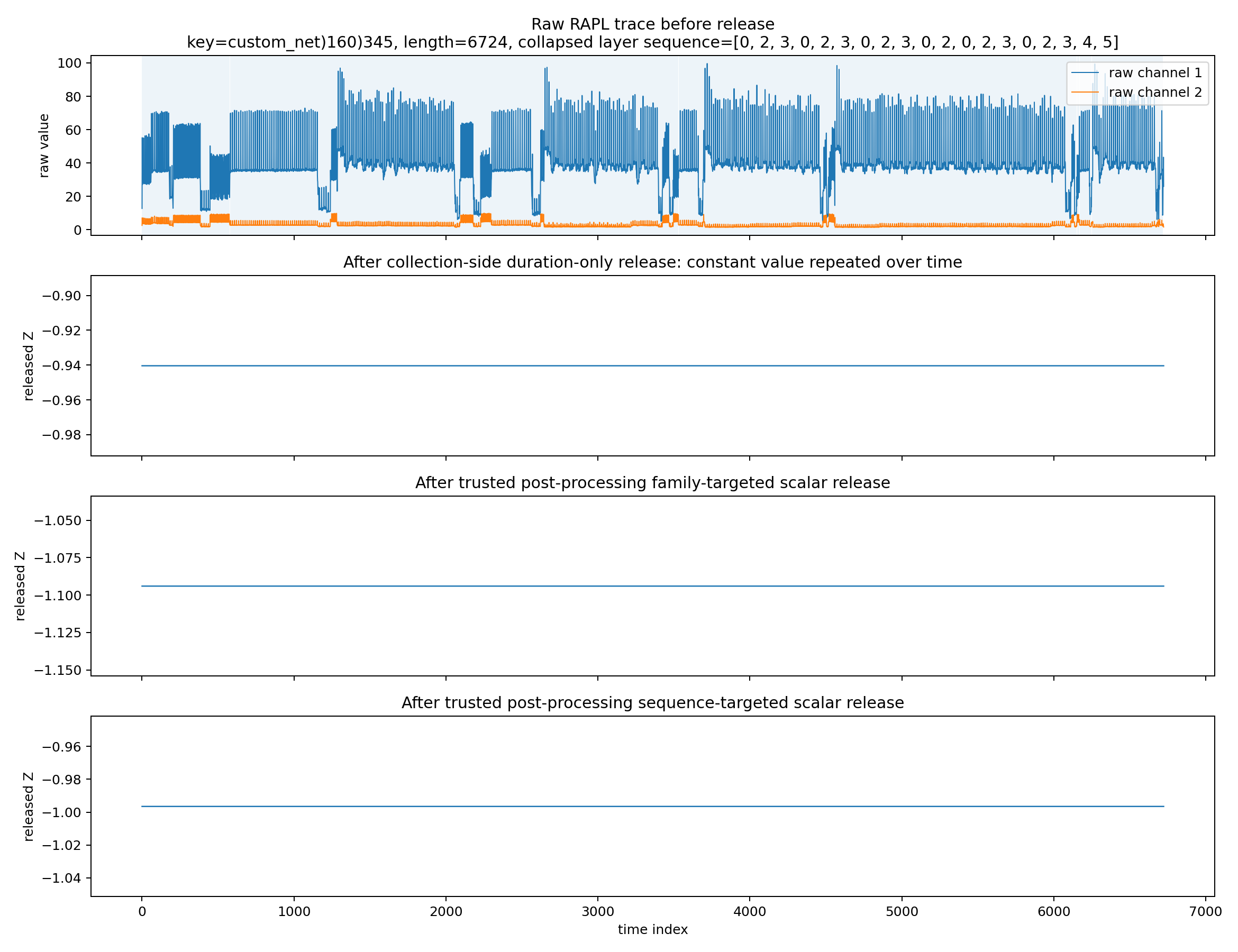}
\caption{One execution from the DeepTheft dataset before and after release design.}
\label{fig:deeptheft-trace}
\end{figure}

The rich baseline is a 256-dimensional downsampled trace. The collection-time alternative is normalized log duration
\[
d_i=\frac{\log(1+T_i)-\mu_T}{\sigma_T},
\]
where $T_i$ is trace length and $\mu_T,\sigma_T$ are fitted on the design split. Candidate scalar releases computed privately after collection use the trace representation together with $d_i$, giving 257 private features. Figure~\ref{fig:deeptheft-trace} makes the disclosure difference concrete: the raw RAPL waveform exposes detailed temporal structure used to recover model architecture, whereas each MID interface releases only one scalar. The horizontal repetition in the figure only visualizes that scalar on the waveform's time axis; the verifier receives one value rather than a time series.

Table~\ref{tab:deeptheft-frontier} compares MID's two deployment paths. Duration can be collected directly and verifies the duration-defined compute tier perfectly while substantially reducing model-family leakage. When that remaining disclosure is unacceptable, a post-collection scalar reduces family inference from $0.7211$ balanced accuracy under raw evidence to $0.1419$, near six-class chance, while retaining $0.9897$ compute-tier accuracy. For architecture, the maximum-utility scalar retains perfect compute-tier verification while reducing conditional architecture inference to $0.2163$, near five-class chance. The result shows when MID's post-collection path adds value: it suppresses information that remains in the best available direct measurement.

\begin{table}[!tbp]
\centering
\caption{DeepTheft matched-support frontier. Sensitive balanced accuracy is averaged within compute tier.}
\label{tab:deeptheft-frontier}
\resizebox{\textwidth}{!}{%
\begin{tabular}{lllccccc}
\toprule
Sensitive target & Evidence & Mechanism & $\beta$ & $\sigma$ & $Y$ BA & $S\mid Y$ BA & $D_S^{\mathrm{cond}}$\\
\midrule
\multirow{4}{*}{Model family (6 classes)}
& Raw features & raw baseline & -- & -- & 0.9099 & 0.7211 & 1.8438\\
& Duration only & collection time & -- & -- & 1.0000 & 0.2929 & 0.0829\\
& Minimum $D_S^{\mathrm{cond}}$ & trusted scalar & 1000 & 0.10 & 0.9897 & 0.1419 & 0.0024\\
& Maximum utility & trusted scalar & 3 & 0 & 0.9989 & 0.2875 & 0.0416\\
\midrule
\multirow{4}{*}{Architecture bucket (5 classes)}
& Raw features & raw baseline & -- & -- & 0.8893 & 0.4754 & 2.6191\\
& Duration only & collection time & -- & -- & 1.0000 & 0.3469 & 0.1113\\
& Minimum $D_S^{\mathrm{cond}}$ & trusted scalar & 10 & 1.00 & 0.8661 & 0.2021 & 0.0167\\
& Maximum utility & trusted scalar & 300 & 0.10 & 1.0000 & 0.2163 & 0.0095\\
\bottomrule
\end{tabular}}
\end{table}
Balanced chance is $1/6\approx0.167$ for model family and $1/5=0.20$ for architecture. The held-out prediction accuracies of the minimum-$D_S^{\mathrm{cond}}$ candidates are $0.1419$ for model family and $0.2021$ for architecture, indicating little conditional discrimination by these models.

\begin{figure}[!tbp]
\centering
\includegraphics[width=0.9\textwidth]{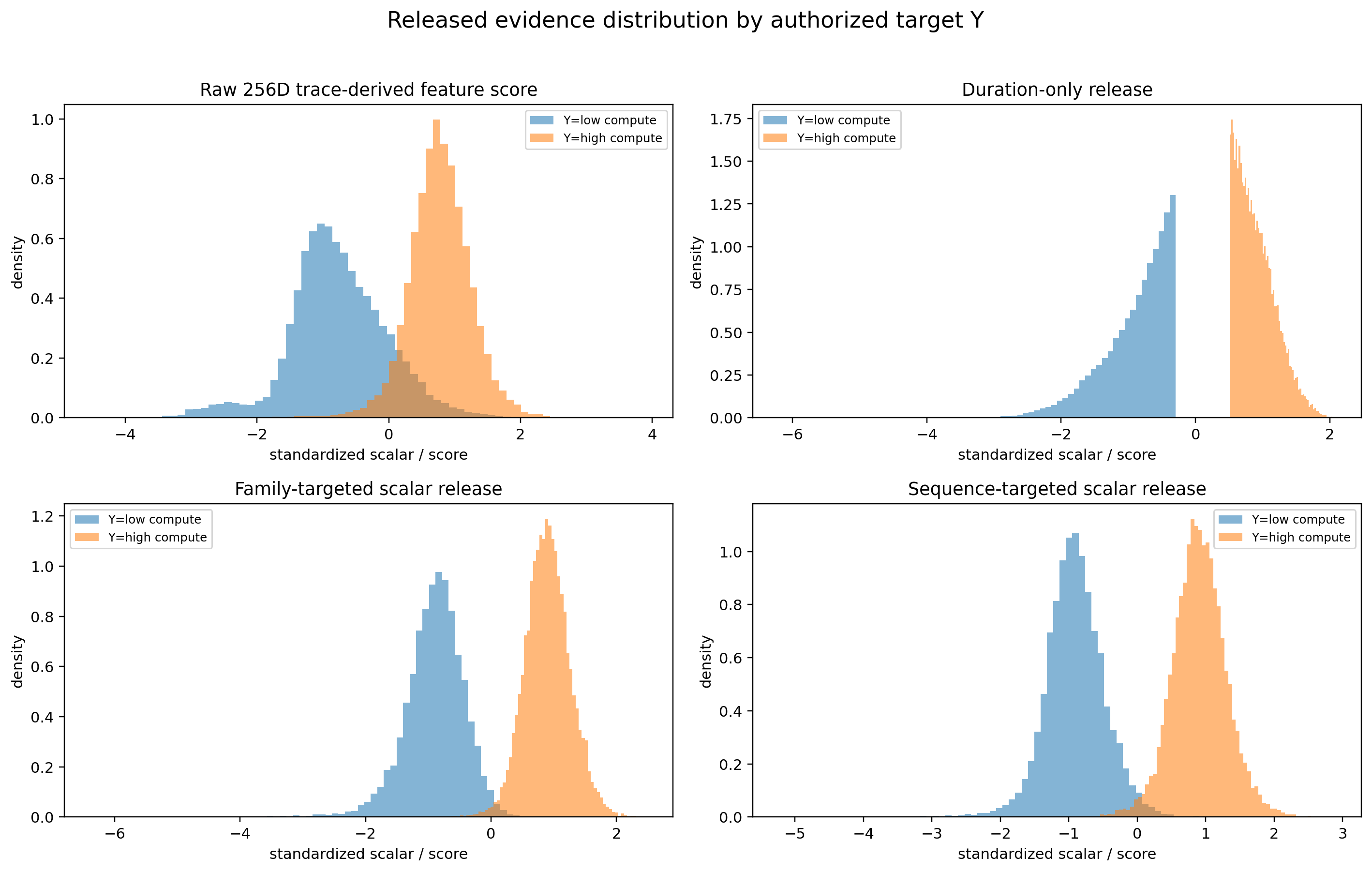}
\caption{DeepTheft evidence grouped by authorized compute tier. The duration and scalar releases preserve the low-versus-high distinction after removing the detailed waveform.}
\label{fig:deeptheft-y-detail}
\end{figure}

\begin{figure}[!tbp]
\centering
\includegraphics[width=\textwidth]{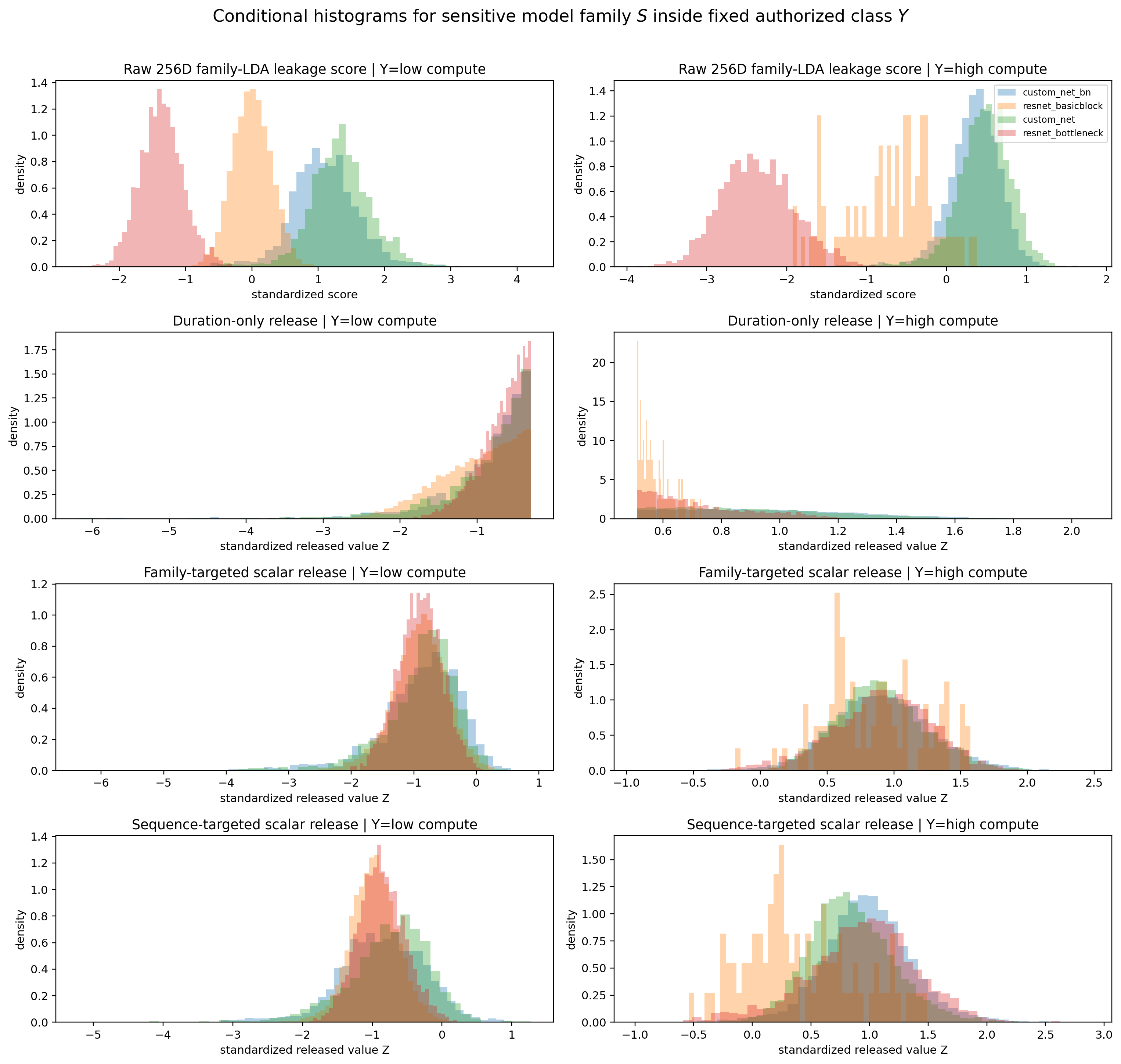}
\caption{DeepTheft model-family evidence within fixed compute tier. Raw trace features contain strong family-separating directions. Duration reduces but does not eliminate that structure; the family-targeted scalar moves conditional inference close to six-class chance. In the plot label, LDA denotes the linear discriminant analysis model used only to create the visualization score.}
\label{fig:deeptheft-s-detail}
\end{figure}

Figures~\ref{fig:deeptheft-y-detail} and~\ref{fig:deeptheft-s-detail} show the two requirements simultaneously. The duration and selected scalar releases retain separation between the authorized low- and high-compute tiers, while the model-family distributions that are clearly separated in the raw evidence become strongly overlapping. MID therefore preserves the coarse verification claim while removing the temporal and distributional structure used to infer the protected model family.

To test whether the low-information releases weaken the DeepTheft attack, we rerun Steps~1 and~2 using Gao et al.'s targets and splits~\cite{deeptheft}. We keep the three-epoch schedule fixed and replace only the attack input. Because the implementation expects a time series, we repeat each scalar release at every time step and add a small $\epsilon$ to UPloss to prevent NaN values. Step~1 recovers the ordered layer-type sequence; we report its Levenshtein Distance Accuracy as the sequence score. Step~2 recovers layer hyperparameters and reports macro F1. Table~\ref{tab:deeptheft-attacks} shows near-perfect recovery from raw RAPL traces. Low-information releases reduce the sequence score to $11.5$--$15.7\%$ and macro F1 to $25.8\%$, with the sequence-targeted scalar performing best on its intended label.

\begin{table}[!tbp]
\centering
\caption{DeepTheft attack evaluation with the attack models, labels, splits, and training schedule held fixed. Replacing only the raw RAPL input with a MID release reduces layer-sequence recovery from $99.1\%$ to $11.5\%$ and macro F1 from $99.8\%$ to $25.8\%$.}
\label{tab:deeptheft-attacks}
\resizebox{\textwidth}{!}{%
\begin{tabular}{llcccc}
\toprule
Attack input & Mechanism type & Epochs & Step~1 sequence score (\%) & Step~2 acc. (\%) & Step~2 F1 (\%)\\
\midrule
Raw RAPL trace & raw baseline & 3 & 99.053 & 99.956 & 99.845\\
Duration only & collection time & 3 & 11.611 & 63.149 & 25.804\\
Family-targeted scalar & trusted post-collection & 3 & 15.668 & 63.149 & 25.804\\
Sequence-targeted scalar & trusted post-collection & 3 & 11.543 & 63.149 & 25.804\\
\bottomrule
\end{tabular}}
\end{table}

Step~2 is class-imbalanced: its $63.149\%$ accuracy under every low-information release accompanies macro F1 $25.804\%$, consistent with majority-class behavior. Macro F1 is therefore the informative Step~2 metric. Because the attack implementation is unchanged and only its evidence is replaced, this collapse shows that MID's disclosure reduction transfers to the original DeepTheft attack, not only to the information estimator used during mechanism selection. This is fundamentally due to MID being an attack-agnostic information-theoretic framework. 

\paragraph{ZKP-backed release.}
We compile the family-targeted scalar into a Groth16 circuit~\cite{groth2016} for Eq.~\eqref{eq:zk-fixed}. The private input contains 256 RAPL features, log duration, and one noise value; the fixed-point scale is 1,000 and $\sigma=0.10$. For the demonstrated instance, $\widetilde\sigma=100$, $\widetilde\xi=126$, $\sum_j\widetilde w_j\widetilde X_j=-2{,}181{,}841$, and $\widetilde Z=-2{,}169{,}241$—Table~\ref{tab:zk-instance} reports its cost. Verification accepts the scalar field-arithmetic relation without revealing $\widetilde X$ or $\widetilde\xi$. 

\begin{table}[!tbp]
\centering
\caption{Groth16 zk-SNARK arithmetic implementation for one scalar report derived from DeepTheft traces.}
\label{tab:zk-instance}
\begin{tabular}{ll@{\qquad}ll}
\toprule
Quantity & Value & Quantity & Value\\
\midrule
Curve & BN254 (\texttt{bn128}) & Circuit constraints & 258\\
Private inputs & 258 & Public inputs & 259\\
Serialized proof & 4 KB & Verification key & 52 KB\\
Proving time & 1.111 s & Verification time & 1.006 s\\
\bottomrule
\end{tabular}
\end{table}

\paragraph{Declared node allocation on NLR.}
\label{sec:nodes}

The NLR dataset combines measured workload profiles with facility power estimated by a bottom-up model of the power-delivery components~\cite{nlr2026power}. Our processed training corpus contains modeled profiles for Llama-2 70B LoRA and Stable Diffusion runs across 2, 4, 8, and 16 nodes. $Y$ is exact node count and $S$ is workload identity within node count. Here, a node means one Kestrel GPU compute server participating in the distributed job—not one GPU. Each Kestrel GPU node contains four NVIDIA H100 GPUs. The raw baseline contains 27 summaries of duration, energy, power levels and quantiles, temporal variation, autocorrelation, and spectral entropy. Workload identity is a meaningful sensitive target because GPU power can vary with input values even when operation shapes are fixed~\cite{gregersen2024power}; an infrastructure signal that verifies aggregate allocation may therefore reveal computation details.

We apply candidate meter reports to $\log_2$ mean power over the following values:
\[
q\in\{0,0.0625,0.125,0.25,0.50,0.75,1.00\},
\qquad
\sigma\in\{0,0.025,0.05,0.10,0.20,0.30,0.40\},
\]
where $q=0$ denotes no quantization and $\sigma$ is the Gaussian-noise standard deviation in log-power units. Selection minimizes conditional workload leakage subject to node-count balanced accuracy of at least $0.90$. The selected deterministic report is
\begin{equation}
Z=\mathrm{round}(\log_2\bar P-b),\qquad b=0.134,
\label{eq:node-report}
\end{equation}
where $\bar P$ is mean facility power. It maps 2, 4, 8, and 16 nodes to codes 12, 13, 14, and 15. Nested four-fold selection chooses $q=1$ and $\sigma=0$ independently in every outer fold using training data only; fitted offsets range from $0.1329$ to $0.1778$. The report is correct on every held-out run, while workload balanced accuracy is $0.5$ within every node count.

\begin{table}[!tbp]
\centering
\caption{NLR node-allocation verification and conditional workload leakage. Mutual information is reported only for the discrete MID channel and is a held-out plug-in estimate.}
\label{tab:node-results}
\begin{tabular}{lcccc}
\toprule
Evidence & $Y$ BA & $S\mid Y$ BA & $I(Y;E)$ & $I(S;E\mid Y)$\\
\midrule
Raw trace features (27) & 0.9773 & 0.9750 & -- & --\\
Raw $\log_2$ mean power & 1.0000 & 0.8729 & -- & --\\
MID discrete report & 1.0000 & 0.5000 & 1.9952 & $1.7\times10^{-16}$\\
\midrule
Balanced chance & 0.2500 & 0.5000 & -- & --\\
\bottomrule
\end{tabular}
\end{table}

Four equally likely node classes carry two bits. For this experiment we instead use the observed run frequencies, whose entropy is $H(Y)=1.9952$ bits. On the held-out runs, the discrete report retains all of it: $I(Y;Z)=1.9952$ bits, while the plug-in estimate $I(S;Z\mid Y)=1.7\times10^{-16}$ bits is numerically zero.

\begin{figure}[!tbp]
\centering
\includegraphics[width=0.95\textwidth]{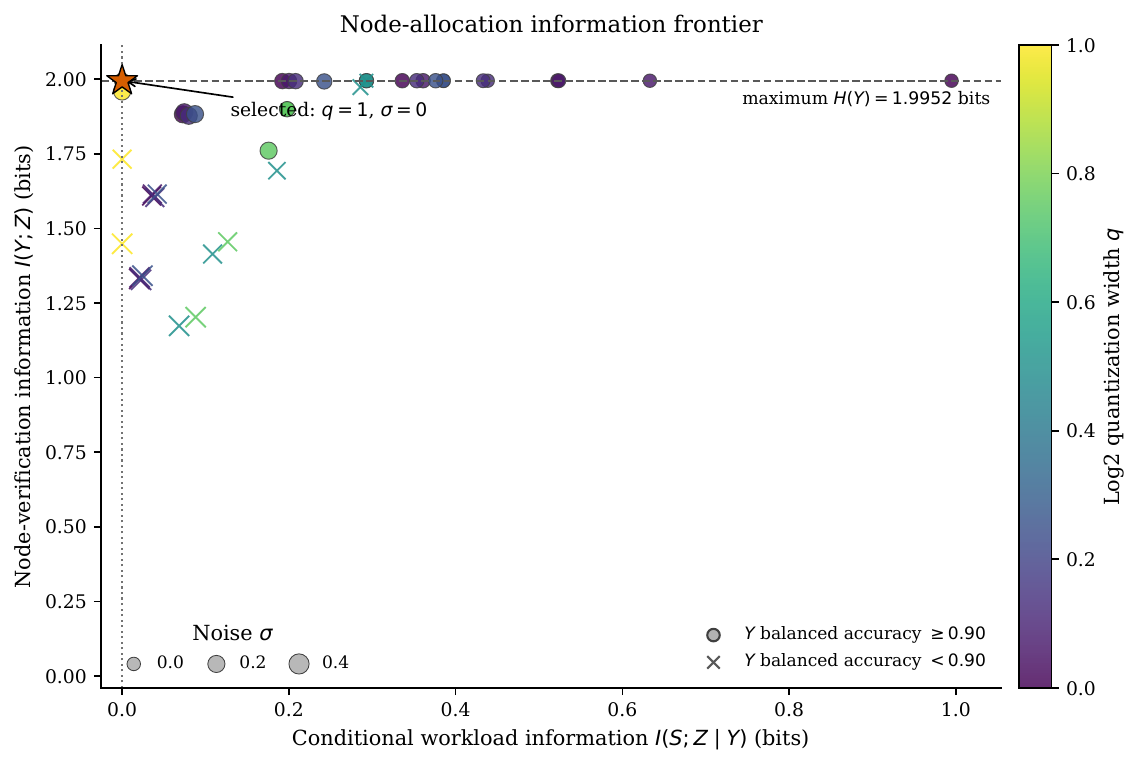}
\caption{NLR node-allocation information frontier. On the held-out runs, the selected quantized report retains all available node-count information and has numerically zero measured conditional workload information.}
\label{fig:node-frontier}
\end{figure}

Figure~\ref{fig:node-frontier} places the selected report on the measured privacy--utility frontier: it retains all available node-count information while reaching numerically zero measured conditional workload information.

When every power value is multiplied by the same calibration factor, the fixed quantizer remains perfect from $-15.5\%$ to $+18.5\%$ error; failures begin at $-16\%$ and $+19\%$. This sweep measures tolerance to uniform calibration error.

\begin{figure}[!tbp]
\centering
\includegraphics[width=0.76\textwidth]{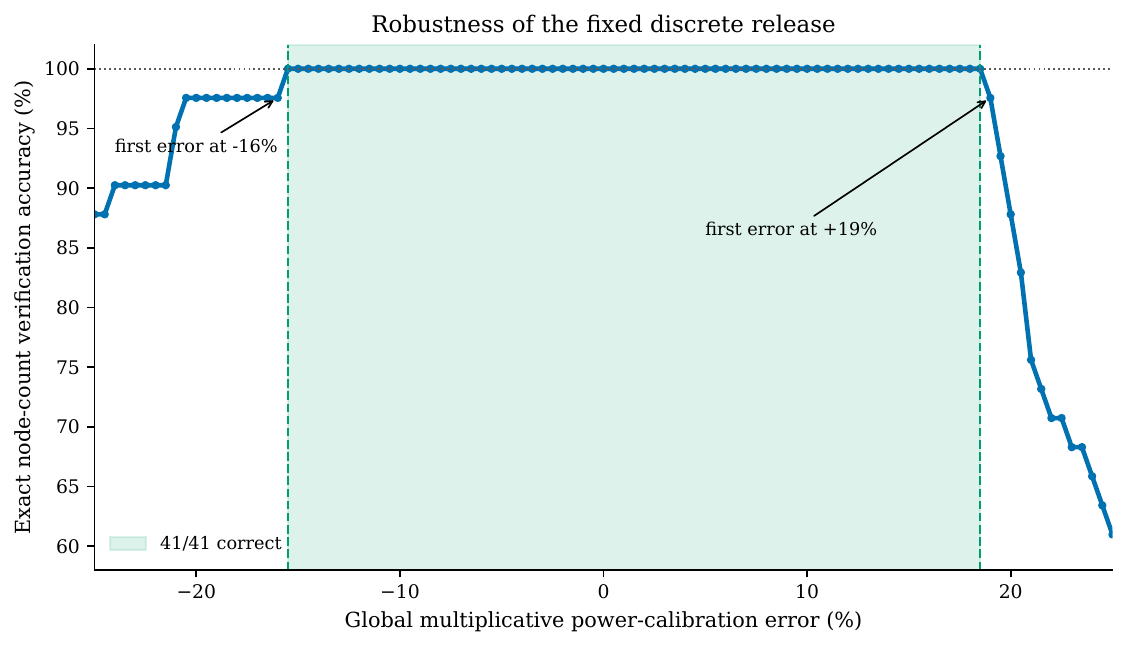}
\caption{Robustness of the selected node-count report to global power-calibration error. Verification remains perfect from $-15.5\%$ to $+18.5\%$ error, showing that the low-information discrete release does not depend on exact power calibration within this interval.}
\label{fig:node-robustness}
\end{figure}

Figure~\ref{fig:node-robustness} visualizes this tolerance: verification remains perfect throughout the stated interval and begins to fail only after the calibration error crosses a quantization boundary.

\begin{figure}[!tbp]
\centering
\includegraphics[width=\textwidth]{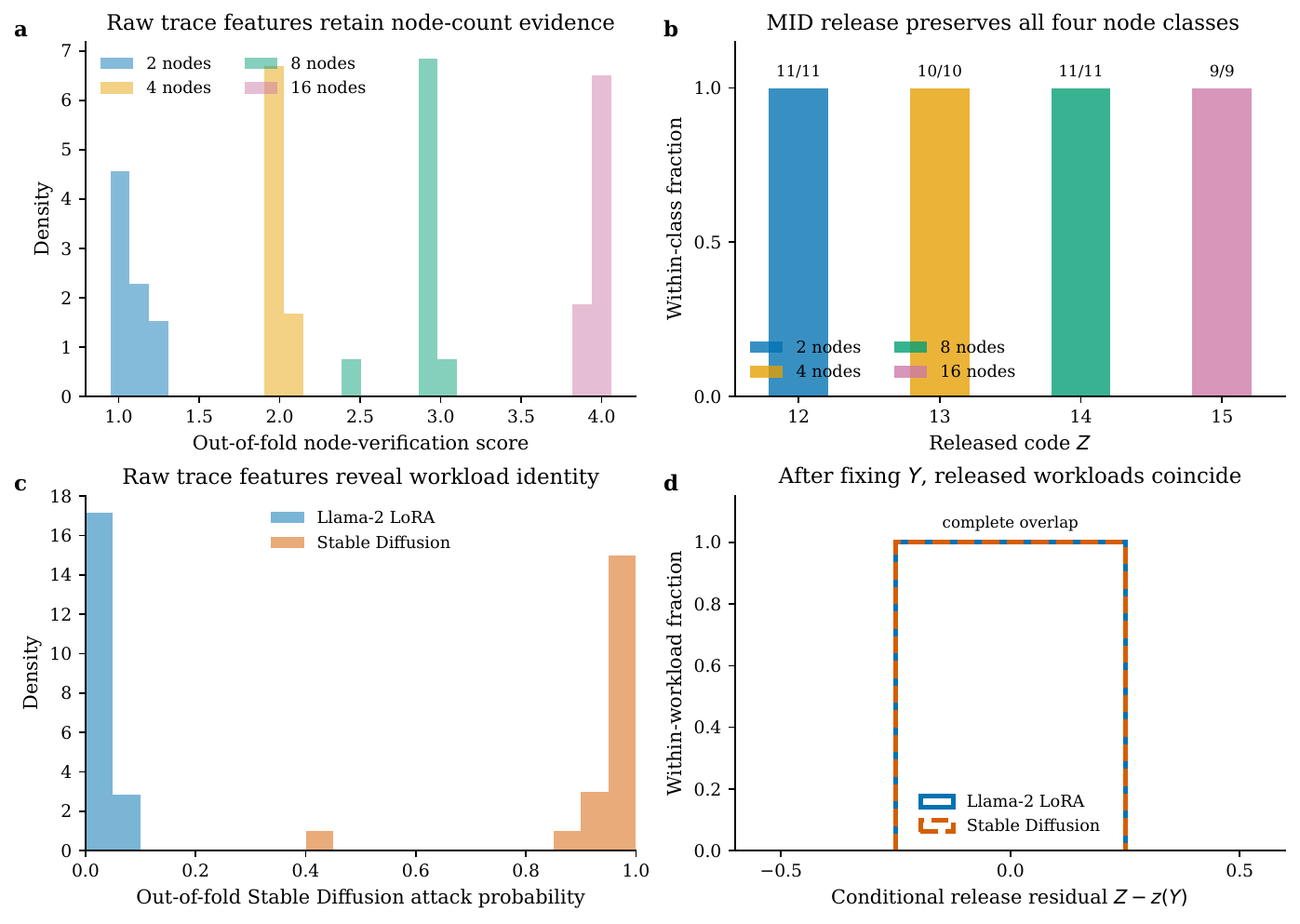}
\caption{NLR node-allocation evidence. The MID-selected report retains four non-overlapping node codes. At each fixed node count, Llama-2 LoRA and Stable Diffusion induce the same released value, hence zero leakage.}
\label{fig:node-detail}
\end{figure}

Figure~\ref{fig:node-detail} shows that the zero-leakage result follows directly from the released code. Each node allocation maps to a distinct value, so the verifier recovers the complete authorized allocation. Within each fixed node count, however, Llama-2 LoRA and Stable Diffusion map to exactly the same value. The release therefore preserves all measured allocation information while providing no measured basis for distinguishing the protected workloads.

\subsection{Model Identity}
\label{sec:modelspy}

Finally, we apply MID to model identity: whether an approved model type executed without exposing its internal structure. Our corpus, filtered from the ModelSpy dataset, contains GPU EM traces from CNN and Transformer executions labeled with model structure~\cite{modelspy}. We authorize $Y\in\{\mathrm{CNN},\mathrm{Transformer}\}$ and protect the finer layer signature $S$, which records the model's ordered layer structure. Candidate collection mechanisms use downsampling factors
\[
\alpha\in\{1,32,48,64,80,96,112,128,160,192,256,320\}.
\]
We simulate lower-rate collection by retaining every $\alpha$-th recorded sample. This assumes the same sensor and analog electronics. In deployment, the sensor would collect at the selected lower rate, so the high-rate digital trace would never be created.

Among candidates with $Y$ balanced accuracy at least $0.90$, $\alpha=192$ minimizes $D_S^{\mathrm{cond}}$. Relative to the raw rate, authorized balanced accuracy changes from $0.9997$ to $0.9046$, $D_S^{\mathrm{cond}}$ falls from $64.2547$ to $7.2474$, and scores from the ModelSpy layer and hyperparameter attacks~\cite{modelspy} fall from $0.9785$ and $0.9468$ to $0.5458$ and $0.5952$, respectively.

\begin{figure}[!tbp]
\centering
\includegraphics[width=\textwidth]{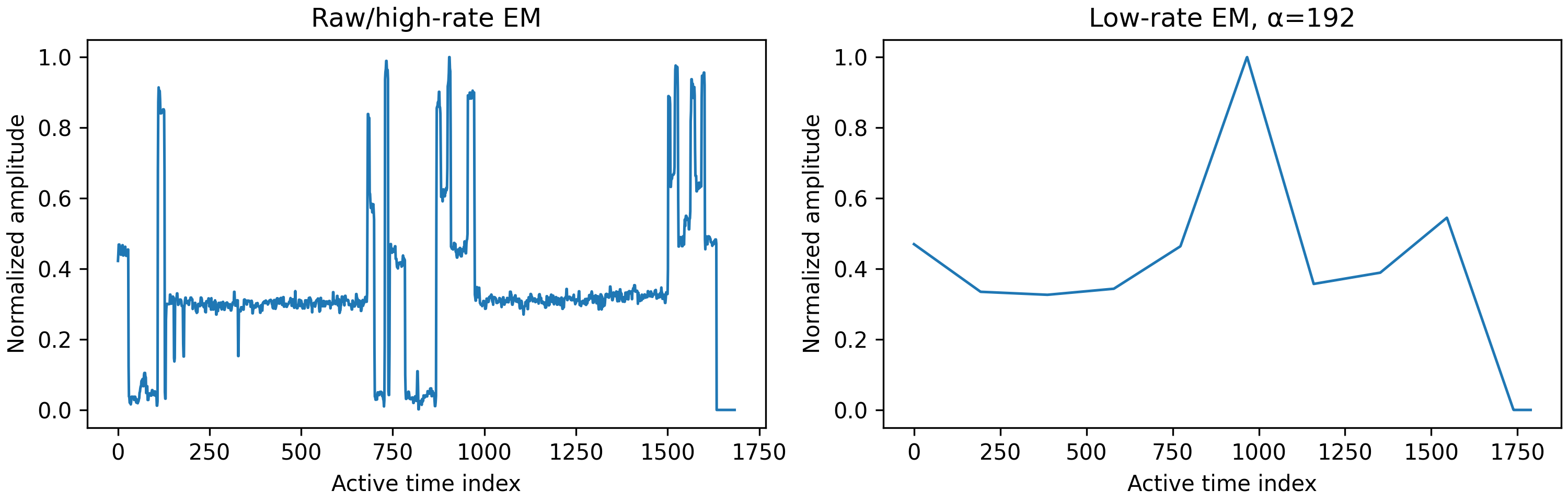}
\caption{Representative ModelSpy EM evidence at the raw rate and selected lower-rate factor $\alpha=192$. The selected collection policy retains the coarse activity pattern used for model-type verification while avoiding most of the fine temporal detail exploited for architecture inference.}
\label{fig:modelspy-trace}
\end{figure}

Figure~\ref{fig:modelspy-trace} shows what MID changes at collection time. The selected lower-rate channel retains the coarse execution pattern needed for CNN-versus-Transformer verification while avoiding most of the fine-grained digital samples available to ModelSpy. This is a privacy advantage of mechanism design at the evidence source rather than post-processing a high-rate trace after it has already been disclosed.

\begin{figure}[!tbp]
\centering
\includegraphics[width=\textwidth]{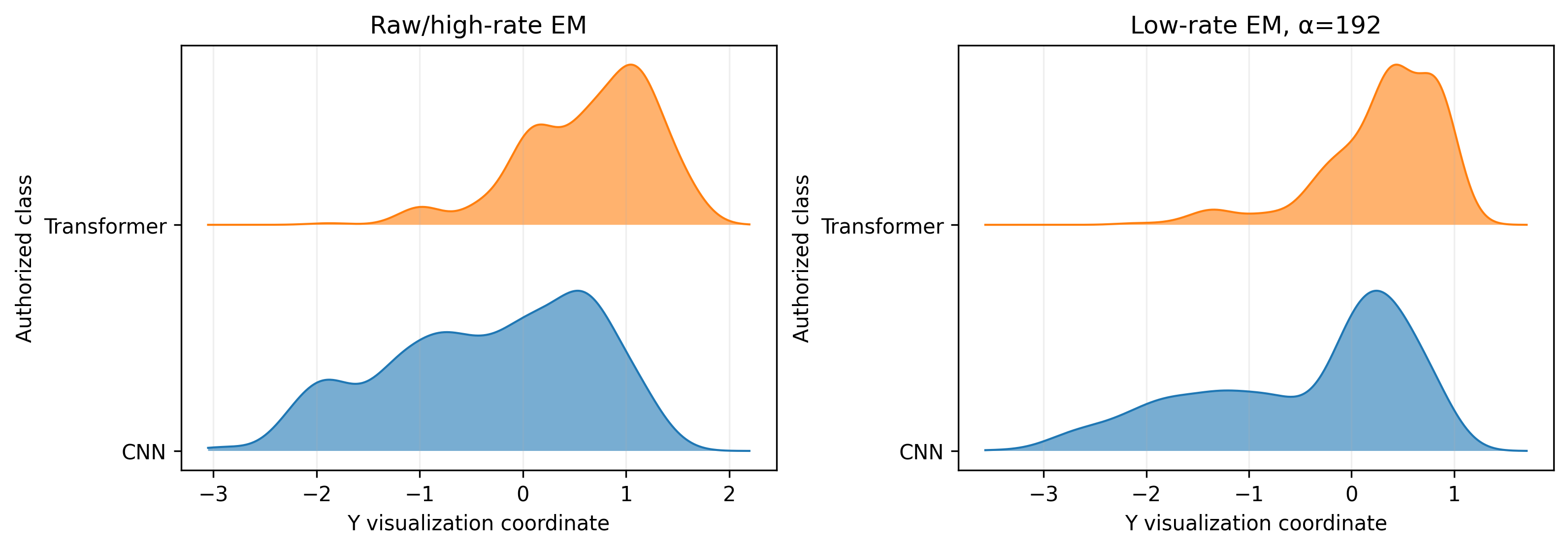}
\caption{Authorized CNN-versus-Transformer evidence before and after lower-rate collection. The selected $\alpha=192$ mechanism retains sufficient class separation to meet the verification requirement, so its privacy improvement is not obtained by making the evidence useless for model-type verification.}
\label{fig:modelspy-y-detail}
\end{figure}

\begin{figure}[!tbp]
\centering
\includegraphics[width=\textwidth]{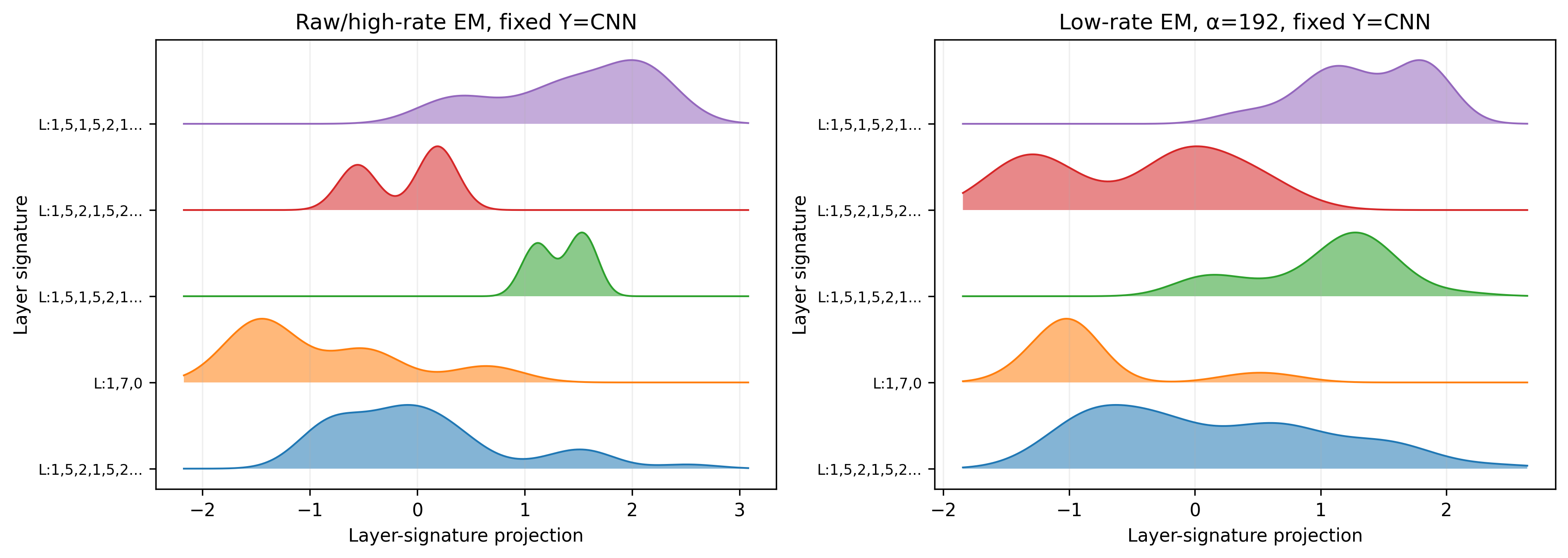}
\caption{Protected layer-signature evidence for the five most frequent layer signatures at fixed $Y=\mathrm{CNN}$. Lower-rate acquisition makes the signature distributions substantially more overlapping, consistent with the reduction in ModelSpy's layer and hyperparameter attack scores.}
\label{fig:modelspy-s-detail}
\end{figure}

Figures~\ref{fig:modelspy-y-detail} and~\ref{fig:modelspy-s-detail} show the resulting tradeoff. At $\alpha=192$, the CNN and Transformer score distributions remain sufficiently separated to satisfy the required $0.90$ authorized balanced accuracy. Within the fixed CNN class, however, the layer-signature distributions overlap substantially more, corresponding to large reductions in the ModelSpy layer and hyperparameter attack scores. MID therefore removes fine model-structure evidence while retaining the coarser model-type distinction.

Table~\ref{tab:modelspy-sweep} gives the complete sweep. Leakage does not decrease smoothly between $\alpha=160$ and $192$. MID therefore selects from the measured frontier rather than assuming that more downsampling always improves privacy.

\begin{table}[!tbp]
\centering
\small
\caption{ModelSpy lower-rate EM sweep. MID selects $\alpha=192$ because it has the lowest measured conditional disclosure among candidates satisfying $Y$ balanced accuracy of at least $0.90$. Candidates with still lower disclosure fail the verification requirement, showing why the mechanism must be selected from the measured privacy--utility information frontier rather than by maximizing downsampling.}
\label{tab:modelspy-sweep}
\resizebox{\textwidth}{!}{%
\begin{tabular}{rcccccc}
\toprule
$\alpha$ & $Y$ acc. & $Y$ bal. acc. & $D_Y$ & $D_S^{\mathrm{cond}}$ & Layer-attack score & Hyperparameter-attack score\\
\midrule
1   & 0.9994 & 0.9997 & 10.4196 & 64.2547 & 0.9785 & 0.9468\\
32  & 0.9889 & 0.9889 & 5.5503  & 42.7931 & 0.7335 & 0.7329\\
48  & 0.9823 & 0.9820 & 5.0442  & 30.6026 & 0.6620 & 0.6905\\
64  & 0.9768 & 0.9848 & 3.0585  & 22.6998 & 0.6408 & 0.6723\\
80  & 0.9453 & 0.9617 & 3.3963  & 17.6194 & 0.6163 & 0.6469\\
96  & 0.9558 & 0.9557 & 2.0929  & 14.9535 & 0.6080 & 0.6421\\
112 & 0.9176 & 0.9440 & 1.7084  & 11.9663 & 0.5865 & 0.6186\\
128 & 0.9022 & 0.9333 & 1.5643  & 11.0527 & 0.5766 & 0.6176\\
160 & 0.8530 & 0.8922 & 1.5446  & 8.0075  & 0.5617 & 0.5941\\
\textbf{192} & \textbf{0.8507} & \textbf{0.9046} & \textbf{1.8943} & \textbf{7.2474} & \textbf{0.5458} & \textbf{0.5952}\\
256 & 0.7584 & 0.8397 & 0.8321  & 5.2068  & 0.4857 & 0.5628\\
320 & 0.6987 & 0.7970 & 0.6889  & 4.0702  & 0.4430 & 0.5337\\
\bottomrule
\end{tabular}}
\end{table}


\section{Discussion}
\label{sec:discussion}

The evaluation shows that MID is not a single release mechanism but a way to choose what evidence should cross the verification boundary under a stated policy. We now discuss the scope of that protection and how the selected interface integrates with existing verification systems.

\paragraph{MID designs what the evidence reveals.}
Across the experiments, MID controls disclosure in different ways: telemetry channels for hardware identity and RL detection, lower-rate acquisition for ModelSpy EM traces~\cite{modelspy}, duration or a private projection for DeepTheft power traces~\cite{deeptheft}, and compact quantized reports for the NLR facility-power data~\cite{nlr2026power}. Searching only over noise added after collection would miss several of these solutions. No construction is universally optimal: the allowed controls and required utility determine the best achievable tradeoff.

\paragraph{What MID protects.}
MID protects a declared property $S$ against disclosure beyond an authorized result $Y$ under the design distribution represented by the development data. Conditional mutual information measures that disclosure; for the corresponding population quantity, the data-processing inequality ensures that later processing of the release cannot increase its information about $S$. A deployment specifies $(Y,S,\tau_Y)$ before mechanism selection, validates on grouped physical runs, and monitors distribution shift. Once policy specifies the authorized claim and protected properties, MID makes the disclosure boundary quantitative and auditable. Differential privacy addresses a different question by bounding worst-case changes under a neighboring-execution relation.

\paragraph{What the experiments show.}
The experiments evaluate the complete mechanism-selection process on held-out physical sessions or runs. In three tasks, the selected discrete report preserves perfect verification and has zero measured conditional leakage on held-out data. The remaining tasks yield explicit privacy--utility frontiers, while the ModelSpy and DeepTheft evaluations show how the selected releases weaken concrete inference attacks~\cite{modelspy,deeptheft}. The same framework can search richer mechanism catalogs and apply Eq.~\eqref{eq:robust-objective} when a deployment requires several protected properties.

\paragraph{Integration with existing verification methods.}
MID complements systems that authenticate evidence, attest execution, or prove computation over private inputs. A deployment first states the authorized result $Y$, the protected property $S$, and the required verification utility. The existing verification system then supplies the mechanisms it can implement with its integrity guarantees: available measurements, collection rules, reports, and required verifier-visible transcript. MID treats each complete verifier-facing interface as $E_m$, including the report and any metadata, attestation, proof, or failure signal the verifier receives, and selects the one that meets the verification requirement while revealing the least about $S$ beyond $Y$. Source authentication or attestation binds evidence to the claimed device or execution; when a release is computed from committed private evidence, a ZKP certifies that computation.

\emph{Protected hardware and confidential execution.}
The flexHEG guarantee processor and Guaranteeable Memory chiplet are designed to observe accelerator activity locally and authenticate detailed receipts or higher-level claims~\cite{petrie2025flexheg,petrie2025guaranteeable}. Attestable Audits runs an agreed benchmark over a confidential model and audit data and publishes an attested aggregate result, while PAL*M attests properties of generative-model operations without exposing the underlying models or datasets~\cite{schnabl2025attestable,chantasantitam2026palm}. MID can compare the measurements, sampling policies, and report functions that each platform can implement. When the selected low-information measurement can be formed at collection time, the monitor need not create or retain a richer trace; otherwise the detailed measurement remains inside the protected boundary and only the selected report and required attestation fields leave it.

\emph{Network taps.}
Cankaya's architecture captures network traffic while keeping its plaintext inside the monitored facility. The verifier receives signed commitments to that traffic and may challenge selected records, which are checked inside an air-gapped auditing environment. The verifier then receives only a report agreed in advance~\cite{cankaya2026system}. MID treats the commitments and report as the verifier-facing evidence and selects the report design that meets the verification requirement while revealing the least about the protected property. If the report exactly communicates the authorized result and the remaining evidence reveals nothing further, the interface is already ideal under MID. The underlying system remains responsible for the integrity of the network tap and auditing process.


\emph{ZKPs and confidential verification.}
ZkAudit proves an agreed audit function over committed private data or model weights, while South et al. prove inference and aggregate evaluation metrics for models with committed private weights~\cite{waiwitlikhit2024zkaudit,south2024zkeval}. Peign\'e et al. propose a dense-training architecture whose public commitments, network anchors, policy bounds, and proof transcript verify a private training specification~\cite{peigne2026zkfrontier}. MID can select among the audit functions, aggregate metrics, or policy claims supported by these protocols and treats every field revealed to the verifier as part of the evidence; the ZKP then proves the selected statement. In our DeepTheft demonstration, a Groth16 zk-SNARK certifies one MID-selected linear projection of a private measurement; in deployment, the surrounding evidence system authenticates the measurement's source~\cite{groth2016}.

\section{Related Work}
\label{sec:related}

MID sits at the intersection of constrained disclosure, confidential verification, evidence generation, and information-theoretic privacy. The central distinction is that MID selects verifier-facing evidence according to what it reveals about a protected property beyond the authorized result.

\paragraph{Privacy requirements in AI verification.}
Prior work constrains what reaches the verifier in several ways: single-bit or yes/no reports~\cite{sastry2024compute,baker2025sixlayers,harack2025verification}, highly specified disclosures~\cite{ogara2025hem}, and on-device aggregation of detailed receipts into higher-level claims~\cite{petrie2025flexheg}. The Agentic Witnessing protocol restricts each answer to four values and budgets the number of queries, yielding a worst-case bound on the capacity of a session~\cite{rowstron2026agentic}. These approaches constrain the form or maximum amount of disclosure, but they do not decide which feasible interface reveals the least about a declared protected property beyond the authorized answer. MID solves that selection problem subject to the required verification utility.

\paragraph{Confidential verification.}
Attestable Audits runs a model and benchmark inside a TEE, publishes an attested aggregate result, and keeps the model weights and audit code and data confidential; PAL*M attests properties of generative-model operations using confidential CPU--GPU execution without exposing the underlying models or datasets~\cite{schnabl2025attestable,chantasantitam2026palm}. ZkAudit proves a verifier-supplied audit function over committed private data and weights, whereas South et al. prove inference and aggregate evaluation metrics for models with committed private weights~\cite{waiwitlikhit2024zkaudit,south2024zkeval}. Peign\'e et al. propose an architecture that combines a precommitted training specification, network observations, Merkle commitments, and ZKPs to verify predeclared training claims~\cite{peigne2026zkfrontier}. These systems protect the inputs to a selected computation but leave the verifier-facing result to the audit designer; MID supplies a quantitative criterion for selecting that result and the mechanism that produces it. Li et al. instead protect individual records with local differential privacy and infer preprocessing errors by comparing black-box model behavior with reference models trained under candidate preprocessing pipelines, explicitly measuring the resulting privacy--verification tradeoff~\cite{li2025ppverification}. MID applies to a different object: the information disclosed by verifier-facing execution evidence about a declared protected property beyond the authorized result.

\paragraph{Evidence generation, soundness, and information flow.}
Proof-of-learning and training-data-verification methods provide evidence about training history or provenance, while subsequent work demonstrates serious soundness weaknesses in current proof-of-learning designs~\cite{jia2021pol,choi2023potd,fang2023polbroken}. Inference-verification methods use recomputation, token divergence, or activation fingerprints~\cite{cankaya2026bitexact,karvonen2025difr}. Guaranteeable Memory places a chiplet at the HBM interface to attest memory snapshots, sampled operations, selected memory regions, or locally checked workload claims; Cankaya instead commits network traffic through signed hashes and evaluates challenged preimages inside an air-gapped auditing environment~\cite{petrie2025guaranteeable,cankaya2026system}. Monfared et al. expose GPU timing and memory telemetry and discuss timing leakage, while Rahman and Tajdari use telemetry to classify hidden training workloads~\cite{monfared2026telemetry,rahman2026hidden}. These methods define candidate evidence channels and integrity paths for MID; MID determines which authenticated interface should reach the verifier. GPU-verification systems also select telemetry features to support reliable verification. WAVE selects performance-counter events to reconstruct model structure and size and proposes a TEE or in-GPU verifier when only the verification result should leave. ShadowScope chooses performance-monitoring events and evaluates how sampling rate affects robust kernel validation~\cite{xu2026wave,almusaddar2025shadowscope}. These systems design evidence for identification and robustness; MID additionally selects it according to what it reveals about a declared $S$ beyond $Y$. Another information-theoretic line in AI verification controls deliberate exfiltration through operational outputs. Rinberg et al. bound steganographic model-weight exfiltration through inference responses. Petrie et al. define unexplained information in a prover's network outputs relative to a declared computation, budget that channel across content, timing, and metadata, and allow the prover to optimize its implementation under that budget~\cite{rinberg2025exfil,petrie2026unexplained}. Those objectives limit information deliberately transmitted through operational outputs. MID measures a different flow: collateral disclosure from legitimate verifier-facing evidence.

\paragraph{General disclosure and information-flow context.}
Outside AI-execution verification, structured transparency distinguishes input from output privacy and calls non-essential information revealed by a flow ``collateral information leakage''~\cite{trask2020structured}. The information bottleneck compresses a representation while preserving task-relevant information~\cite{tishby1999}. The privacy funnel and subsequent privacy--utility work optimize release mechanisms to suppress designated sensitive variables under utility and observation constraints~\cite{makhdoumi2014,sankar2013,wang2017,liao2019}. These works provide the general privacy and information-theoretic context; MID formulates the evidence-design problem for AI verification and links the selected release to an authenticated direct measurement or a ZKP-certified private computation. The breadth of inference attacks on model interfaces and physical traces motivates this information objective rather than a separate defense for each attack~\cite{shokri2017mia,yeom2018privacy,nasr2019whitebox,carlini2022lira,song2021systematic,modelspy,deeptheft,magneticnn,horvath2024sok,gregersen2024power,debenedetti2024sidechannels}.

\section{Conclusion}

AI verification must reveal enough to establish a claim without exposing unnecessary details about the execution. MID turns this tension into a quantitative design problem. Given an authorized target $Y$, a protected property $S$, and a verification requirement $\tau_Y$, MID searches across evidence channels, collection policies, and release transformations for an interface that preserves verification while minimizing collateral leakage. Across six physical-measurement tasks, this process selects restricted channels, lower-rate collection, discrete reports, or private projections and reports the resulting privacy--utility frontiers. The selected interface can be deployed through an authenticated direct measurement or a ZKP-backed transformation of a private measurement; we demonstrate the scalar-release arithmetic with a Groth16 zk-SNARK. MID thus makes what verification evidence reveals an explicit, testable design choice.

\appendix

\section{Output-Privacy Analysis of Auditor-in-a-Box}
\label{app:auditor-output-privacy}

Penchas et al. propose a two-party protocol in which both parties sign a Plan that an LLM executes inside an attested TEE~\cite{penchas2026enabling}. The authors explicitly present a deterministic output filter intended to impose a single-bit worst-case disclosure bound~\cite{rinberg2026auditor}. The reference implementation's Monitor Query Validation example applies this idea through a restricted verdict interface.\footnote{\href{https://github.com/RoyRin/auditor-in-a-TEE/blob/33d8d065510ea1144371085cc0c0515fc8ed1f44/cli/examples/B.1/plan.yaml}{Pinned Monitor Query Validation source.}} In the published evaluation analyzed below, every reported decision is \textsc{valid} or \textsc{invalid}. This limits how much leaves the protected system, but does not determine what the observed verdict reveals.


We test this distinction using only the authors' published evaluation of the usage-monitor classifier.\footnote{\href{https://github.com/RoyRin/auditor-in-a-TEE-webapp/blob/26353c4a307ca38fa851b3ca3530758ecb11029a/roc-eval.html}{Pinned published evaluation.}} We analyze only requests that human reviewers marked \textsc{valid}, so the intended answer is the same for every case. The authors' case identifiers begin with either B or V. Many B-case descriptions call the request adjacent, proxy, or borderline, while V cases largely describe ordinary aggregate-research tasks. Although all of these requests have the same human judgment, the monitor returns \textsc{invalid} for 47.4\% of B cases and only 2.5\% of V cases. The one-bit decision therefore reveals information about what kind of private request was submitted.

Formally, $Y$ is the human judgment, fixed here at \textsc{valid}; $Z$ is the monitor's released \textsc{valid}/\textsc{invalid} decision; and $S$ records whether the case identifier begins with B or V. For each case, a simple predictor learns from all the other cases which prefix is more likely after each decision. It then receives only the decision for the case being tested and predicts B or V; it does not receive the request text, identifier, or purpose. The case identifiers, descriptions, human judgments, and model decisions all come from the authors' evaluation; we add no labels or examples.

\begin{table}[h]
\centering
\caption{Disclosure and decision-only inference at the fixed authorized answer $Y=\textsc{valid}$. Although the interface releases only one bit, $0.176$ bits of measured disclosure increases inference of the protected request group from $50.0\%$ to $72.4\%$ balanced accuracy.}
\label{tab:auditor-output-privacy}
\small
\begin{tabular}{lcc}
\toprule
Metric & Authorized answer only & With binary decision $Z$ \\
\midrule
Measured disclosure beyond $Y$ & 0 & 0.176 bits \\
Estimated B-group probability & 19.4\% & 81.8\% after \textsc{invalid} \\
Inference-attack balanced accuracy & 50.0\% & 72.4\% \\
\bottomrule
\end{tabular}
\end{table}

The important result is not merely that the decision carries $0.176$ bits. It is that this apparently small disclosure supports a decision-only inference attack with $72.4\%$ balanced accuracy, compared with $50.0\%$ without the decision. The experiment therefore demonstrates the central motivation for MID on an existing verification interface: limiting output cardinality bounds how much can be released, but does not determine what the released bit reveals. 


\section*{Acknowledgments}

This work was supported by
\raisebox{-0.28\height}{\includegraphics[height=1.15em]{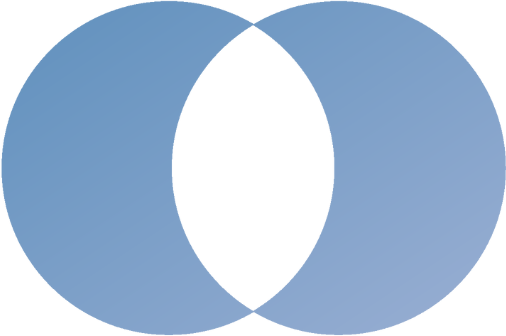}}\,
\href{https://www.pivotal-research.org/}{\textcolor{pivotalblue}{\textbf{Pivotal Research}}}.

\end{document}